\documentclass[trackchanges,twocolumn,twocolappendix]{aastex701}

\usepackage{amsmath}
\usepackage{color}
\usepackage[dvipsnames]{xcolor}
\usepackage{graphicx}
\usepackage{txfonts}
\usepackage{float}
\usepackage[version=4]{mhchem}
\usepackage{booktabs}
\usepackage{soul} 
\usepackage{multirow}
\usepackage{subcaption}
\usepackage{comment}
\usepackage{placeins}
\usepackage{makecell}

\usepackage[T1]{fontenc}
\usepackage{textcomp}

\usepackage{pifont} 
\newcommand{\cmark}{\ding{52}}
\newcommand{\xmark}{\ding{56}}
\newcommand{\qmark}{\textbf{?}}
\newcommand{\emark}{\textbf{!}}

\received{2026 March 30}
\revised{2026 June 2}
\accepted{2026 June 8}

\begin{document}

\title{Probing the ubiquity of complex ices in protostars with JWST:\\ the first systematic quantification of weak ice bands between 6.8 and 7.9~µm}

\author[orcid=0000-0002-3395-5634, sname='Chen', gname='Yuan']{Yuan Chen}
\affiliation{Leiden Observatory, Leiden University, P.O. Box 9513, 2300~RA Leiden, The Netherlands}
\affiliation{Max Planck Institut für Extraterrestrische Physik (MPE), Giessenbachstrasse 1, 85748 Garching, Germany}
\email[show]{yuanchen@mpe.mpg.de, ychen@strw.leidenuniv.nl}

\author[orcid=0000-0001-7591-1907, sname='van Dishoeck', gname='Ewine']{Ewine F. van Dishoeck}
\affiliation{Leiden Observatory, Leiden University, P.O. Box 9513, 2300~RA Leiden, The Netherlands}
\affiliation{Max Planck Institut für Extraterrestrische Physik (MPE), Giessenbachstrasse 1, 85748 Garching, Germany}
\email{ewine@strw.leidenuniv.nl}

\author[orcid=0000-0002-7433-1035, sname=Slavicinska, gname=Katerina]{Katerina Slavicinska}
\affiliation{Laboratory for Astrophysics, Leiden Observatory, Leiden University, PO Box 9513, 2300~RA Leiden, The Netherlands}
\affiliation{Leiden Observatory, Leiden University, P.O. Box 9513, 2300~RA Leiden, The Netherlands}
\email{slavicinska@strw.leidenuniv.nl}

\author[orcid=0000-0002-9470-2358, sname=Tychoniec, gname=Łukasz]{Łukasz Tychoniec}
\affiliation{Leiden Observatory, Leiden University, P.O. Box 9513, 2300~RA Leiden, The Netherlands}
\email{tychoniec@strw.leidenuniv.nl}

\author[orcid=0000-0002-6312-8525, sname=van Gelder, gname=Martijn]{Martijn L. van Gelder}
\affiliation{Leiden Observatory, Leiden University, P.O. Box 9513, 2300~RA Leiden, The Netherlands}
\email{ml.vgelder@gmail.com}

\author[orcid=0000-0001-8822-6327, sname='Francis', gname='Logan']{Logan Francis}
\affiliation{Leiden Observatory, Leiden University, P.O. Box 9513, 2300~RA Leiden, The Netherlands}
\email{francis@strw.leidenuniv.nl}

\author[orcid=0000-0001-9344-0096, sname=Boogert, gname=Adwin]{Adwin C. A. Boogert}
\affiliation{Institute for Astronomy, University of Hawaii at Manoa, 2680 Woodlawn Drive, Honolulu, HI 96822, USA}
\email{aboogert@hawaii.edu}

\author[orcid=0000-0002-0554-1151]{Mayank Narang}
\affiliation{Jet Propulsion Laboratory, California Institute of Technology, 4800 Oak Grove Drive, Pasadena, CA 91109, USA}
\email{mayankn1154@gmail.com}

\author{Charlotte A. F. Schravendijk}
\affiliation{Laboratory for Astrophysics, Leiden Observatory, Leiden University, PO Box 9513, 2300~RA Leiden, The Netherlands}
\email{caf.schravendijk@gmail.com}

\author[orcid=0000-0002-1700-090X, sname=Beuther, gname=Henrik]{Henrik Beuther}
\affiliation{Max Planck Institute for Astronomy (MPIA), Königstuhl 17, 69117 Heidelberg, Germany}
\email{beuther@mpia.de}

\author[orcid=0000-0001-8876-6614]{Alessio Caratti o Garatti}
\affiliation{INAF--Osservatorio Astronomico di Capodimonte, Salita Moiariello 16, 80131 Napoli, Italy}
\affiliation{Dublin Institute for Advanced Studies, 31 Fitzwilliam Place, D02 XF86 Dublin, Ireland}
\email{alessio.caratti@inaf.it}

\author{Eyze K. H. Eenhorst}
\affiliation{Laboratory for Astrophysics, Leiden Observatory, Leiden University, PO Box 9513, 2300~RA Leiden, The Netherlands}
\email{ekh.eenhorst@gmail.com}

\author[orcid=0000-0002-8120-1765, sname=Gieser, gname=Caroline]{Caroline Gieser}
\affiliation{Max Planck Institute for Astronomy (MPIA), Königstuhl 17, 69117 Heidelberg, Germany}
\email{gieser@mpia.de}

\author[orcid=0000-0002-1493-300X, sname=Henning, gname=Thomas]{Thomas Henning}
\affiliation{Max Planck Institute for Astronomy (MPIA), Königstuhl 17, 69117 Heidelberg, Germany}
\email{henning@mpia.de}

\author[orcid=0000-0001-9818-0588]{Manuel Güdel}
\affiliation{Department of Astrophysics, University of Vienna, T\"urkenschanzstrasse 17, A-1180 Vienna, Austria}
\affiliation{ETH Z\"urich, Institute for Particle Physics and Astrophysics, Wolfgang-Pauli-Strasse 27, 8093 Z\"urich, Switzerland}
\email{manuel.guedel@univie.ac.at}

\author[orcid=0000-0002-5714-799X]{Valentin J. M. Le Gouellec}
\affiliation{Institut de Cienciès de l’Espai (ICE-CSIC), Campus UAB, Carrer de Can Magrans S/N, E-08193 Cerdanyola del Vallès, Spain}
\affiliation{Institut d’Estudis Espacials de Catalunya (IEEC), c/ Gran Capitá, 2-4, 08034 Barcelona, Spain}
\email{valentin.le.gouellec@gmail.com}

\author[orcid=0000-0003-1878-327X]{Melissa K. McClure}
\affiliation{Leiden Observatory, Leiden University, P.O. Box 9513, 2300~RA Leiden, The Netherlands}
\email{mcclure@strw.leidenuniv.nl}

\author[orcid=0000-0002-4448-3871]{Pooneh Nazari}
\affiliation{European Southern Observatory (ESO), Karl-Schwarzschild-Strasse 2, Garching, 85748, Germany}
\email{pooneh.nazari@eso.org}

\author[orcid=0000-0003-0276-5368]{Simon D. Reyes}
\affiliation{Max Planck Institute for Astronomy (MPIA), Königstuhl 17, 69117 Heidelberg, Germany}
\email{reyes@mpia.de}

\author{Mees Tolman}
\affiliation{Laboratory for Astrophysics, Leiden Observatory, Leiden University, PO Box 9513, 2300~RA Leiden, The Netherlands}
\email{n401276@gymnasiumnovum.nl}

\author[orcid=0000-0001-8227-2816]{Yao-Lun Yang}
\affiliation{Star and Planet Formation Laboratory, Pioneering Research Institute, RIKEN, 2-1 Hirosawa, Wako, Saitama, 351-0198, Japan}
\email{yaolunyang.astro@gmail.com}


\begin{abstract}
Complex organic molecules (COMs) are the key to understanding the chemical evolution from simple interstellar molecules to potential prebiotic material. Although COMs have been extensively studied in the gas phase toward protostars, their counterparts in ices, where they are thought to form at earlier stages, remain far less constrained. A number of diagnostic features of complex ices lie between 6.8 and 8.8 µm, a region known as the ``COM ice fingerprint range,'' but previous infrared facilities lacked the sensitivity and spectral resolution required to quantify the weak bands therein. With the unprecedented sensitivity and resolving power of JWST, these limitations can now be overcome. Here, we present the first large-sample quantitative study of the absorption features at 7.02, 7.24, 7.40, and 7.67 µm, using MIRI-MRS spectra of 21 protostars. 
The \ce{CH4} band at 7.67 µm is the strongest band and shows remarkably uniform peak positions (7.67–7.68 µm) and FWHMs (0.06–0.08 µm), suggesting \ce{CH4} ice as its dominant carrier. The 7.24 and 7.40 µm bands exhibit larger source-to-source variations in peak positions and FWHMs, but their occurrence and intensities are strongly correlated with each other. Comparisons with existing and new laboratory spectra suggest \ce{HCOO-} as the most likely carrier of these two bands, yet \ce{HCOO-} cannot fully reproduce their intensity ratios, implying additional contributions from other species such as \ce{C2H5OH}, \ce{CH3CHO}, and \ce{CH3COCH3}.
Our results reveal, for the first time, the potential ubiquity of weak features of complex ices in protostars, which have remained largely undetected due to observational limitations.
\end{abstract}



\section{Introduction}\label{sect:introduction}
Understanding the chemical evolution of the Universe from simple atoms in diffuse clouds to the complex chemistry necessary for life is one of the overarching goals of astrochemistry. Complex organic molecules (COMs), typically defined as carbon-containing species with six or more atoms \citep{Herbst2009}, are a key piece in this puzzle due to their potential links to prebiotic chemistry. 
Gas-phase COMs have been detected in various astrophysical environments, but mostly in the warm ($\gtrsim100$~K) inner regions of low- and high-mass protostars, often referred to as hot corinos or hot cores \citep{Jorgensen2020, Ceccarelli2023}. Observations with (sub)millimeter facilities such as the Atacama Large Millimeter/submillimeter Array (ALMA) have revealed rich gas-phase COMs in protostellar sources, indicating a high degree of chemical complexity at early stages of star formation (see the review by \citealt{Nazari2025} and references therein).

The formation history of COMs remains a topic of active research. A growing body of observational, theoretical, and experimental work supports a primarily grain-surface origin for these species. In cold prestellar cores ($T\lesssim10$~K), CO gas freezes out onto dust grains and undergoes successive hydrogenation, forming \ce{CH3OH} and more complex species within the ice mantles \citep[e.g.,][]{Watanabe2002, Fuchs2009, Simons2020, Fedoseev2022}. As the protostar forms and heats its surroundings, these COMs are thermally desorbed into the gas phase and become observable at millimeter (mm) wavelengths through their rotational transitions. However, gas-phase formation routes may also play a role in COM formation \citep{Balucani2015}, especially in the warm-up phase or in regions where energetic processing (e.g., UV irradiation and cosmic rays) is significant \citep{Garrod2022}. The relative contributions of solid-phase inheritance and gas-phase reprocessing are still debated and likely vary among sources and molecules (e.g., \citealt{Nazari2022_NCOM, Chen2023, Chen2024}).

To trace the formation and chemical evolution of COMs, it is crucial to probe their solid-phase reservoirs on dust grains. Infrared spectroscopy provides the only direct probe of interstellar ices by detecting their vibrational bands in absorption. Over the past three decades, observations with IR facilities such as the Infrared Space Observatory (ISO) and the Spitzer Space Telescope have established the dominant ice inventory in protostellar environments (e.g., \citealt{Gibb2004, Boogert2008, Oberg2011, Boogert2015}). However, these observations primarily focused on the strongest absorption bands of silicates and abundant species (e.g., \ce{H2O}, \ce{CO}, and \ce{CO2}) due to limited sensitivity and spectral resolution. Weaker absorption features expected from molecules that are more complex than \ce{CH3OH} have remained much less explored. In particular, the wavelength range between 6.8 and 8.8 µm (often referred to as the ``COM ice fingerprint range'') contains multiple weak and partially blended bands associated with COMs and formate ions (\ce{HCOO-}). Although some of these features have been noted in previous studies (e.g., \citealt{Schutte1999}), it has been very difficult to robustly detect them, let alone quantify their spectral properties across a large sample.

The unprecedented sensitivity and resolving power of the James Webb Space Telescope (JWST) now make it possible to overcome these limitations. Thanks to the Medium Resolution Spectroscopy (MRS) mode of JWST's Mid-InfraRed Instrument (MIRI), absorption features of multiple oxygen-bearing COMs (\ce{CH3CHO}, \ce{C2H5OH}, \ce{CH3OCHO}, and \ce{CH3COCH3}) have been firmly or tentatively detected in mid-IR spectra of a few protostars by focusing on the COM ice fingerprint range (\citealt{Chen2024, Rocha2024, Rayalacheruvu2025, Gieser2026, Turner2026}). The simplest nitrogen-bearing COM, \ce{CH3CN}, has also been tentatively detected using the Near Infrared Spectrograph (NIRSpec; \citealt{Nazari2024_CH3CNice}). These observations provide, for the first time, the opportunity to directly compare the abundances of COMs in both the solid and the gas phases in the same protostellar sources (i.e., B1-c and NGC~1333~IRAS~2A, hereafter IRAS~2A; \citealt{Chen2024}). Such comparisons suggest that both inheritance and reprocessing are plausible when COMs sublimate from icy mantles to the gas phase.

To extend previous studies of COM ice features from individual sources to a larger sample, we present the first systematic quantification of weak absorption bands in the COM ice fingerprint range, using JWST/MIRI-MRS spectra of 21 protostars selected from three programs (see Sect.~\ref{sect:observations}). Rather than focusing on individual detections, we measure and compare the band properties of the absorption features near 7.02, 7.24, 7.40, and 7.67 µm across the sample. This approach allows us to investigate the correlations among the bands and constrain their likely molecular carriers, without performing dedicated fitting with laboratory spectra. By enabling systematic and quantitative characterization of weak absorption bands associated with complex ices, JWST not only opens a new window for studying the origin, evolution, and ubiquity of COMs, but also provides a more complete view of interstellar ice chemistry.


\section{Observations}\label{sect:observations}
The JWST/MIRI-MRS spectra used in this work were collected from three programs focusing on (primarily low-mass) protostars, including two Guaranteed Time Observation (GTO) programs 1290 \citep[JWST Observations of Young protoStars  (JOYS); PI: E. F. van Dishoeck;][]{vD2025_JOYS} and 1236 (JOYS+; PI: M. E. Ressler; M. E. Ressler et al. in preparation), and one General Observers (GO) program 1802 \citep[Investigating Protostellar Accretion (IPA); PI: T. Megeath;][]{Federman2024_IPA}. 
The data reduction of GTO~1290, GTO~1236, and GO~1802 has been described in \cite{vanGelder2024_gas}, \cite{vanGelder2024_SO2}, and \cite{Federman2024_IPA}, respectively.

\begin{table*}[!htb]
    \renewcommand{\arraystretch}{1}
    \centering
    \caption{Source Information and Detection Results of the Four Absorption Bands Studied between 6.8 and 7.9~µm}
    \begin{tabular}{llcccc|cccccccc}
    \hline\hline
    Source & Other Name & {Program}$^\text{a}$ & $D$ & {Class} & {Aperture}$^\text{b}$ & \multicolumn{2}{c}{7.67\,µm} & \multicolumn{2}{c}{7.24\,µm} & \multicolumn{2}{c}{7.40\,µm} & \multicolumn{2}{c}{7.02\,µm} \\ \addlinespace[-1.5pt]
     & & & (pc) & & & {$\sigma^\text{c}$} & & {$\sigma^\text{c}$} & & {$\sigma^\text{c}$} & & {$\sigma^\text{c}$} & \\
    \hline 
    B1-b & Per-emb 41 & JOYS & 293 & I & 4$\times$PSF & 36 & \cmark & 6.9 & \cmark & 6.8 & \cmark & 4.5 & \cmark\emark \\ 
    B1-c & Per-emb 29 & JOYS & 293 & 0 & 4$\times$PSF & 251 & \cmark & 51 & \cmark & 54 & \cmark & 32 & \cmark \\ 
    NGC 1333 IRAS 1A & Per-emb 35A & JOYS+ & 293 & I & 3$\times$PSF & 67 & \cmark & 18 & \cmark & 11 & \cmark & 6.6 & \cmark \\ 
    NGC 1333 IRAS 2A & Per-emb 27 & JOYS+ &  293 & 0 & 4$\times$PSF & 57 & \cmark & 12 & \cmark & 12 & \cmark & 4.6 & \cmark\emark \\ 
    L1448-mm & Per-emb 26 & JOYS & 293 & 0 & 4$\times$PSF & 29 & \cmark & 9.0 & \cmark & 5.2 & \cmark & ... & \xmark \\ 
    L1448-IRS2B & Per-emb 22B & JOYS+ & 293 & 0 & 4$\times$PSF & 90 & \cmark & 24 & \cmark & 20 & \cmark & 10 & \cmark \\ 
    L1448-IRS3B1 & Per-emb 33B & JOYS+ & 293 & 0 & 3$\times$PSF & 64 & \cmark & 11 & \cmark & 12 & \cmark\qmark & 5.1 & \cmark \\ 
    Per-emb 8 &   & JOYS & 321 & 0 & 4$\times$PSF & 7.3 & \cmark & ... & \qmark & ... & \qmark & ... & \qmark \\ 
    \hline
    L1527 & IRAS 04368+2557 & JOYS & 142 & 0/I & 3$\arcsec$\,{\small(fixed)} & 36 & \cmark & 4.3 & \cmark\emark & 4.9 & \cmark\emark & 8.2 & \cmark \\ 
    \hline
    Ser-S68N-N &   & JOYS & 436 & 0 & 3$\times$PSF & 24 & \cmark & 7.6 & \cmark & 4.8 & \cmark\emark & 2.8 & \emark \\ 
    Ser-SMM1A &   & JOYS & 436 & 0 & 3$\times$PSF & 21 & \cmark & 2.3 & \emark & 3.3 & \cmark\emark & ... & \xmark \\
    Ser-SMM1B &   & JOYS & 436 & 0 & 3$\times$PSF & 175 & \cmark & 24 & \cmark & 22 & \cmark & 11 & \cmark \\ 
    Ser-SMM3 &   & JOYS & 436 & 0 & 4$\times$PSF & 12 & \cmark & 2.3 & \emark & 2.8 & \emark & 1.4 & \emark \\ 
    SVS4-5 &   & JOYS & 436 & I/II & 4$\times$PSF & 34 & \cmark & 17 & \cmark & 16 & \cmark & 13 & \cmark \\ 
    (Ser-SMM4$^\text{d}$) &  & & & (0$^\text{d}$)  & & & & & & & & & \\
    \hline
    HOPS 153 &   & IPA & 390 & 0 & 4$\times$PSF & 45 & \cmark & 6.3 & \cmark & 6.2 & \cmark & 7.2 & \cmark \\ 
    HOPS 370 &   & IPA & 390 & 0/I & 7$\times$PSF & 129 & \cmark & 122 & \cmark & 86 & \cmark & 40 & \cmark \\ 
    \hline
    BHR71-IRS1 & IRAS 11590--6452 & JOYS & 200 & 0 & 4$\times$PSF & 166 & \cmark & 34 & \cmark & 28 & \cmark & 23 & \cmark \\ 
    BHR71-IRS2 &   & JOYS & 200 & 0 & 4$\times$PSF & 15 & \cmark & 2.4 & \emark & 2.1 & \emark & ... & \qmark \\ 
    \hline
    B335 & IRAS~19345+0727 & IPA & 165 & 0 & 4$\times$PSF & 53 & \cmark & 7.9 & \cmark & 6.3 & \cmark & ... & \xmark \\ 
    IRAS~16253-2429 &   & IPA & 140 & 0 & 4$\times$PSF & 63 & \cmark & 9.0 & \cmark & 5.5 & \cmark & 8.2 & \cmark \\ 
    IRAS~20126+4104 &   & IPA & 1550 & 0 & 4$\times$PSF & 219 & \cmark & 38 & \cmark & 26 & \cmark & 18 & \cmark \\ 
    \hline
    \end{tabular}
    
    \vspace{2pt}

    \begin{minipage}{0.92\textwidth}
    $^\text{a}$ JOYS = GTO~1290, JOYS+ = GTO~1236, IPA = GO~1802.\\
    $^\text{b}$ The extraction aperture in diameter. PSF is an abbreviation of FWHM$_\text{PSF}=0.033\arcsec\times(\lambda/\text{µm})+0.106\arcsec$, where $\lambda$ is the median wavelength of each MIRI-MRS Channel (i.e., the extraction aperture increases with wavelength).\\
    $^\text{c}$ Indication of different symbols: \cmark~($\geq5\sigma$; robust detection), \cmark\emark~($<5\sigma$ and $\geq3\sigma$; detection), \emark~($<3\sigma$ and $\geq1\sigma$; tentative detection), \qmark~(this band is visible but has an irregular shape), \xmark~(this band is not detected or not recognizable due to a low SNR or contamination of gas-phase lines).\\
    $^\text{d}$ SVS4-5 is a Class I/II source located behind the envelope of the Class 0 source Ser-SMM4 in the line of sight.\\
    \end{minipage}
    \label{tab:source_info}
\end{table*}

A total of 21 sources with at least one absorption band (i.e., the \ce{CH4} band at 7.67~µm) robustly detected in the COM ice fingerprint range were selected from the three programs (see Table~\ref{tab:source_info}), of which 12 are from GTO 1290 (JOYS), four from GTO 1236 (JOYS+), and five from GO 1802 (IPA). 
We excluded less embedded later-stage (Class I/II) sources that show barely detected ice features or a flat spectral energy distribution due to their lower envelope masses, and those that are too faint (embedded) to have enough signal-to-noise ratios (SNRs) for absorption features. 
Except for one intermediate-mass (HOPS~370) and one high-mass (IRAS~20126) source, all the other sources considered in this work are low-mass. Although the JOYS program contains several high-mass sources, their COM ice fingerprints have been presented in other studies (\citealt{Rocha2024, vD2025_JOYS, Gieser2026}); therefore, we only include the low-mass sample in JOYS. For the IPA sample, since no dedicated study of COM ices has been carried out, we include all the five sources even though they cover a wide range of luminosities (0.16--$10^4\,L_\odot$; \citealt{Federman2024_IPA}). 

All the spectra were extracted from the IR continuum peak positions to maximize the SNRs of absorption features. The extraction aperture in diameter was usually set to 3--4 times the point spread function (PSF) of MIRI-MRS, which is given by 
$\text{FWHM}_\text{PSF} = 0.033\arcsec\times(\lambda/\mathrm{\mu m}) + 0.106\arcsec$ \citep{Law2023}. 
A slightly different aperture was used to include extended emission in some sources (e.g., the scattered light of L1527; see Table~\ref{tab:source_info}). We refer to \cite{vD2025_JOYS} and \cite{Federman2024_IPA} for other detailed information about the sources from JOYS and IPA. An overview paper of JOYS+ sources is upcoming (M. E. Ressler et al. in preparation).

\section{Methods}\label{sect:methods}
In order to quantitatively analyze the weak absorption bands in the COM ice fingerprint range, we need to remove any contaminating gas-phase lines (especially the rovibrational lines of gas-phase \ce{SO2} at $\sim$7.35~µm and \ce{CH4} at 7.66~µm) and trace a local continuum ($F_\text{local cont}$) to isolate the spectra in the relevant wavelength range; meanwhile, the observed spectra in the flux scale ($F_\text{obs}$) are converted into the optical depth scale via $\tau_\text{obs}=-\ln(F_\text{obs}/F_\text{local cont})$. We give the details of these two steps in Appendix~\ref{appendix:methods}.

There are two major differences between the analysis of the COM ice fingerprint range in this work and the previous case studies by \cite{Rocha2024} and \cite{Chen2024}. First, we do not investigate the full 6.8--8.8~µm region in this work, but only focus on the ``blue'' half of it (i.e., 6.8--7.9~µm; shortward of the \ce{CH4} band at 7.67~µm). This is because the absorption features beyond $\sim$7.9~µm are generally weak and have irregular shapes (see Fig.~\ref{fig:full_range}). The broad and deep silicate band at 9.8~µm lowers the flux, and hence the SNR in this range. It is also intrinsically difficult to determine where to place the local continuum without performing a global fitting for all the absorption features relevant to this wavelength range (e.g., \citealt{Rayalacheruvu2025, Turner2026}). As a result, the characterization of absorption features in 7.9--8.8~µm is much less reliable than that in 6.8--7.9~µm (see Appendix~\ref{appendix:7.9-8.8um} for further discussion).
Second, we do not aim to perform dedicated fitting with laboratory spectra and derive ice column densities of \ce{H2O} and other molecules. Instead, we measure the properties such as peak wavelengths and full widths at half-maximum (FWHMs) of four absorption bands observed between 7.0 and 7.7~µm, and compare these values with those measured from laboratory spectra of synthetic ices (see Sect.~\ref{sect:results_peak_width}), which is only possible with large sample analysis.

\begin{figure}[!htb]
    \centering
    \includegraphics[width=\linewidth]{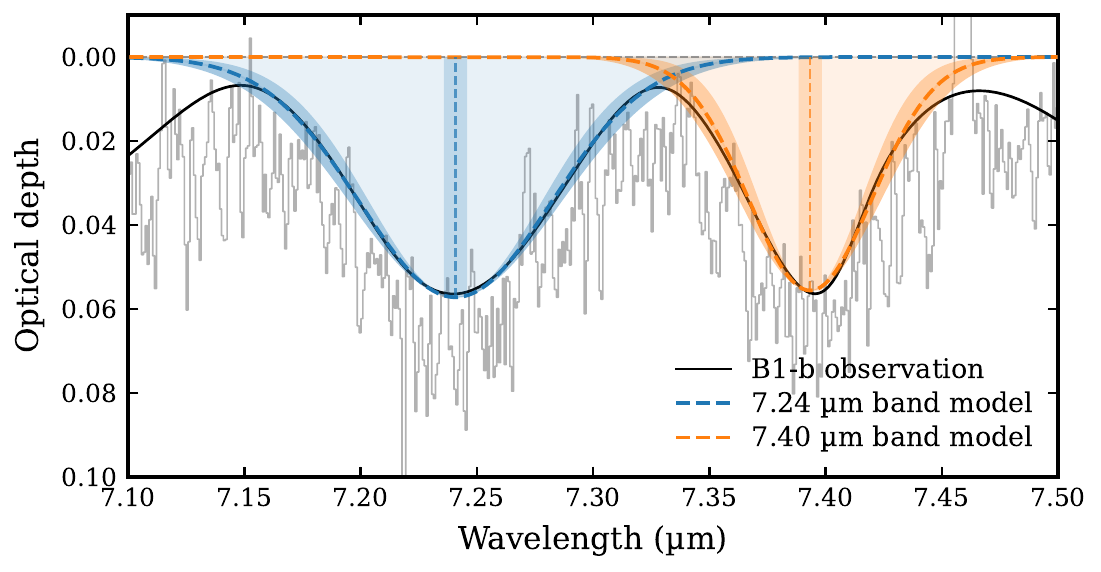}
    \caption{Gaussian fitting of the 7.24 and 7.40~µm bands, taking B1-b as an example. The observations before and after gas-line removal are shown as thin gray and thick black lines, respectively. 
    The shaded regions in blue and orange represent the two Gaussian functions jointly fitted to the 7.24 and 7.40~µm bands, respectively, with the vertical dashed lines marking the centers of the fitted Gaussian components}. The darker shaded regions indicate the uncertainties in the peak wavelength (0.005~µm) and FWHM (0.01~µm); see Appendix~\ref{appendix:uncertainty} for further description of the uncertainty estimation.
    \label{fig:72-74um_fit}
\end{figure}

After isolating the weak absorption features, we measure four properties, peak wavelength ($\lambda_\text{peak}$), peak optical depth ($\tau_\text{peak}$), FWHM, and integrated area ($I_\text{band}$), for the four absorption bands at 7.02, 7.24, 7.40, and 7.67~µm, which have been attributed to \ce{CH3CHO}, \ce{C2H5OH}, \ce{HCOO^-}, and \ce{CH4} ices.
The first three properties ($\lambda_\text{peak}$, $\tau_\text{peak}$, and FWHM) were either measured by Gaussian fitting (e.g., Fig.~\ref{fig:72-74um_fit}) or by definition (i.e., the wavelength at which the spectral feature reaches its maximum optical depth, and the width of the feature measured at half of its peak value), depending on whether the observed band shapes can be well reproduced by Gaussian functions or not.
The integrated area of a band is calculated by $I_\text{band} = \int_{\tilde{\nu}_1}^{\tilde{\nu}_2} \tau_{\tilde{\nu}}\ \text{d}\tilde{\nu}$,
where $\tilde{\nu}$ is the wavenumber, and $\tilde{\nu}_1$ and $\tilde{\nu}_2$ mark the band range.
The fitting results of band properties and the fitting strategies (i.e., by Gaussian or by definition) are listed in Table~\ref{tab:band_prop}. We adopt a uniform and conservative estimation for the uncertainties of $\lambda_\text{peak}$ (0.005~µm), FWHM (0.01~µm), and $I_\text{band}$ (20\%). More details on the uncertainty estimation are provided in Appendix~\ref{appendix:uncertainty}.

We also estimate noise levels in optical depth ($\tau_\sigma$; listed in Table~\ref{tab:band_prop}) around the four studied absorption bands, using channels free from emission or absorption features in the spectra before removing any gas-phase lines. The SNR in multiples of $\sigma$ is calculated for each absorption band by dividing its $\tau_\text{peak}$ by $\tau_\sigma$ (listed in Table~\ref{tab:source_info}), providing a quantified detection significance of each band.

\section{Results and discussion}\label{sect:results_discussion}
\subsection{Detection of ice bands}\label{sect:results_detection}

\begin{figure}[!htb]
    \centering
    \includegraphics[width=\linewidth]{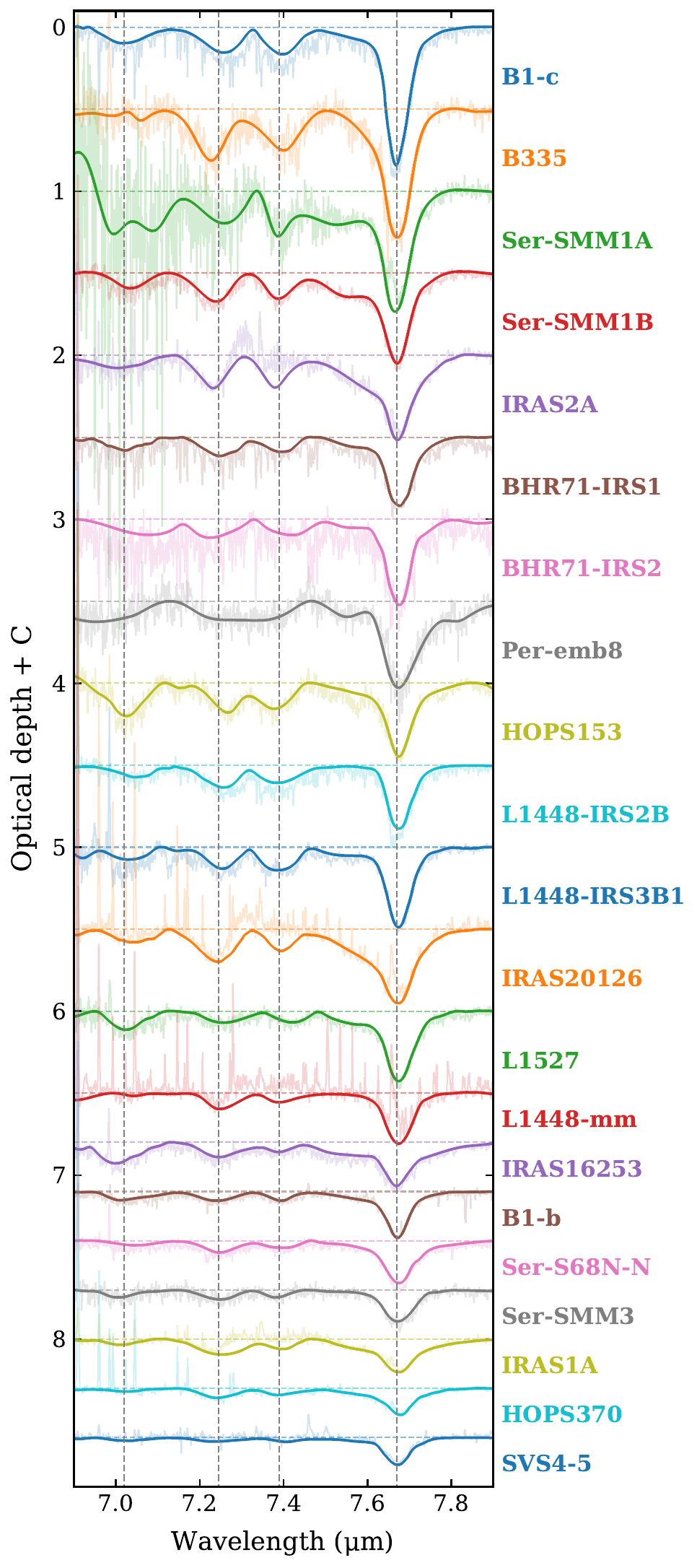}
    \caption{Isolated absorption features between 6.9 and 7.9~µm after subtracting the local continuum (and removing the rovibrational lines of gas-phase \ce{SO2} and \ce{CH4}, if necessary). The spectra are displayed from top to bottom in a rough order of the intensity of the \ce{CH4} band at 7.67~µm. The spectra plotted in thin and thick lines are the spectra before and after removing the gas-phase lines, respectively. Four vertical dotted lines in gray are plotted at 7.02, 7.245, 7.39, and 7.67~µm, respectively, for comparison with the observed absorption bands.}
    \label{fig:overview_COM_range}
\end{figure}

Figure~\ref{fig:overview_COM_range} presents the isolated spectra between 6.9 and 7.9~µm (i.e., the blue half of the COM ice fingerprint range) of the 21 selected protostellar sources. 
As expected from the selection criterion, the \ce{CH4} band at 7.67~µm is robustly detected in all of them. The two bands at 7.24 and 7.40~µm, which have been attributed to \ce{HCOO^-}, \ce{C2H5OH}, or \ce{CH3CHO} ices, are detected together and show Gaussian-like shapes in the majority of the sample, except for Ser-SMM1A, Ser-SMM3, BHR71-IRS2, and Per-emb~8, where the detection is considered tentative due to low SNRs (see Table~\ref{tab:source_info}). The co-occurrence of the 7.24 and 7.40~µm bands in observations provides an important clue to the identification of \ce{HCOO-} ice, which has two bands at $\sim$7.24 and 7.39~µm. A more quantitative analysis is presented in Sect.~\ref{sect:results_area}.

The 7.02~µm band is located in a fainter range of the spectrum, on top of the red wing of the strong \ce{NH4^+} band at 6.85~µm; therefore, its shape and intensity are more susceptible to the local continuum than the other three bands mentioned above. The 7.02~µm band can be attributed to \ce{CH3OH}, \ce{CH3CHO}, \ce{C2H5OH}, and \ce{CH3COCH3}. Its detection is robust for high-SNR sources such as B1-c, BHR71-IRS1, HOPS~370, and IRAS~20126, but in general, its shape is less regular, and its SNRs are lower than those of the bands at 7.24 and 7.40~µm.

\begin{figure*}[!htb]
    \centering
    \includegraphics[width=\textwidth]{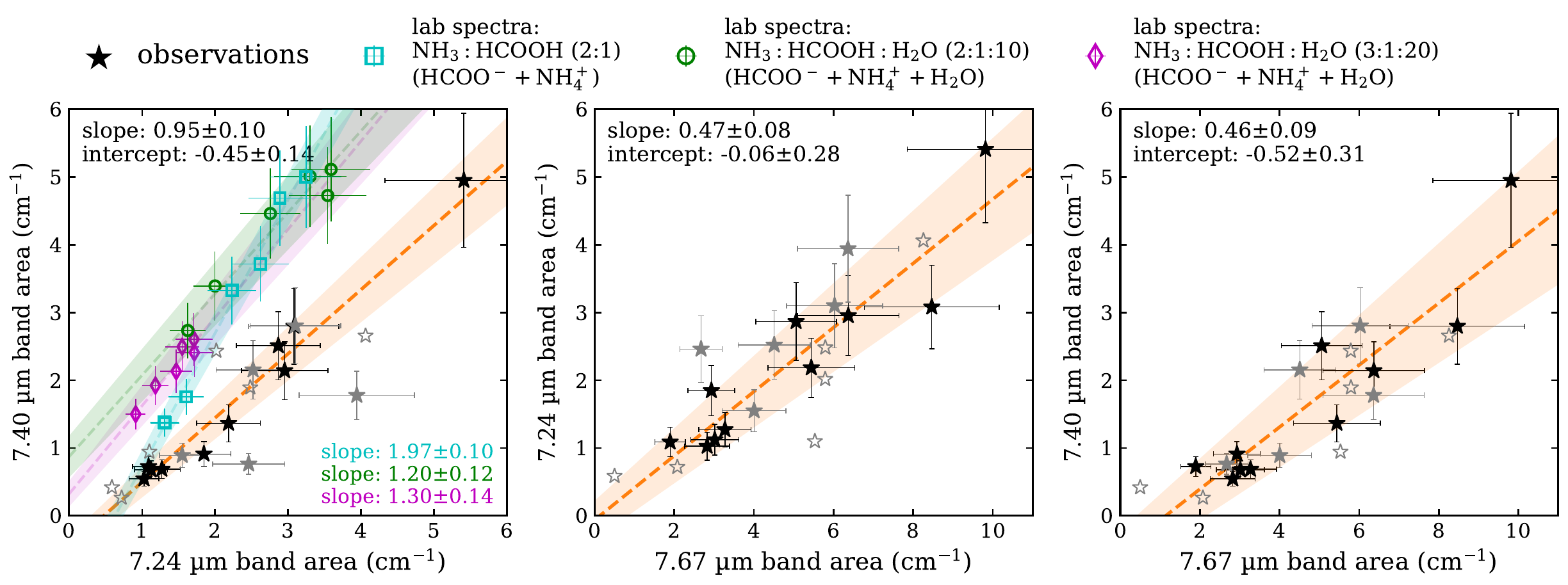}
    \caption{Integrated areas of the three bands at 7.24, 7.4, and 7.67~µm. Values measured from observations are denoted by stellar markers in black or gray; markers in hollow gray and solid gray indicate the observed 7.24 or 7.40~µm bands are in low SNRs and contaminated by the gas-phase lines, respectively. Only the solid points (i.e., those with a robust SNR) were taken into account in the fitting. In each panel, the best linear fit of the observations and its uncertainty are indicated by the dashed line and shaded region in orange, respectively, with the value of the best-fit slope and intercept annotated in the upper left corner. In the left panel, areas of two \ce{HCOO^-} bands at 7.24 and 7.39~µm measured from laboratory spectra at different temperatures (between 15 and $\sim$170~K, see Table~\ref{tab:HCOO-_band_prop}) are added for comparison with the observations. Cyan, purple, and green correspond to three \ce{HCOO^-} ices: the pure one and two \ce{H2O}-mixed ones with different mixing ratios (see the common legend at the top). The shaded regions in these three colors indicate the ranges of laboratory-measured band area ratios of different \ce{HCOO^-} ices, and the slope of the best linear fits is annotated in the lower right corner. The experimental band areas of pure \ce{HCOO^-} ices are scaled by a factor of 2/3 to facilitate the comparison with observational results. 
    The band area uncertainties for the observed and laboratory spectra are set as 20\% and 15\%, respectively (see the explanations in Appendix~\ref{appendix:uncertainty}).}
    \label{fig:band_ratios}
\end{figure*}

\subsection{Band properties I: integrated area}\label{sect:results_area}
The integrated area of an absorption band is related to the ice column densities of contributing species by $N_\text{ice}=I_\text{band}/A$, where $A$ is the band strength (in units of cm~molecule$^{-1}$) measured by experiments. 
Figure~\ref{fig:band_ratios} displays the correlations between the $I_\text{band}$ of the three bands at 7.24, 7.40, and 7.67~µm. The linear correlations are fitted with a Markov Chain Monte Carlo, 
with sources in low SNRs excluded. The integrated areas of the three bands are strongly correlated with each other:
\begin{align}
    I_{7.24~\mu\text{m}} &= (0.95\pm0.10)\ I_{7.40~\mu\text{m}} - (0.45\pm0.14);\\
    I_{7.24~\mu\text{m}} &= (0.47\pm0.08)\ I_{7.67~\mu\text{m}} - (0.06\pm0.28);\\
    I_{7.40~\mu\text{m}} &= (0.46\pm0.09)\ I_{7.67~\mu\text{m}} - (0.52\pm0.31).
\end{align}
These correlations strengthen the qualitative conclusion about the co-occurrence of the 7.24 and 7.40~µm bands made in Sect.~\ref{sect:results_detection}; they also suggest that nondetection of weak absorption bands is more likely caused by limited sensitivity or SNRs, rather than the absence of ice reservoirs in the observed protostars (e.g., \citealt{Rocha2025}). 

The co-occurrence of the observed bands at 7.24 and 7.40~µm also provides support for the long-standing attribution to \ce{HCOO^-} ice, which has two bands at $\sim$7.24 and 7.39~µm \citep{Schutte1999}. Formate ion (\ce{HCOO^-}) as a salt in interstellar ices is considered a key tracer of acid–base reactions and energetic processing, both of which are important and efficient pathways that transform simple molecules into more complex ones in ices under astrochemical environments. To investigate if \ce{HCOO^-} ice is the only carrier of the two bands at 7.24 and 7.40~µm, we take a step further and compare their areas in observations with those measured in the new laboratory spectra of \ce{HCOO^-} ice. In laboratories, \ce{HCOO^-} ice is produced through acid-base reactions in HCOOH:\ce{NH3} mixtures (see Appendix~\ref{appendix:HCOO-_lab_spec} for more details). Here, we considered pure and two \ce{H2O} mixtures of \ce{HCOO^-} ices, and measured the properties ($\lambda_\text{peak}$, FWHM, and $I_\text{band}$) of their 7.24 and 7.39~µm bands from IR spectra (listed in Table~\ref{tab:HCOO-_band_prop}). The left panel of Fig.~\ref{fig:band_ratios} shows that the 7.24~µm band of \ce{HCOO^-} ice in laboratory spectra is generally weaker than its 7.39~µm band, with the slope of $I_{7.24~\mu\text{m, HCOO}^-}/I_{7.39~\mu\text{m, HCOO}^-} = 1.20$--1.97 ($>$1.0), but in observations, the 7.24~µm band is slightly stronger (slope of $I_{7.24~\mu\text{m, obs}}/I_{7.40~\mu\text{m, obs}}<1.0$). This difference in band area ratios provides a strong evidence that \ce{HCOO^-} may not be the only carrier of the observed bands at 7.24 and 7.40~µm, and there are likely other species contributing to the 7.24~µm band (e.g., \ce{C2H5OH}, see Sect.~\ref{sect:results_peak_width}).



\subsection{Band properties II: peak and width}\label{sect:results_peak_width}
\begin{figure*}[!hp]
    \centering
    \includegraphics[width=\textwidth]{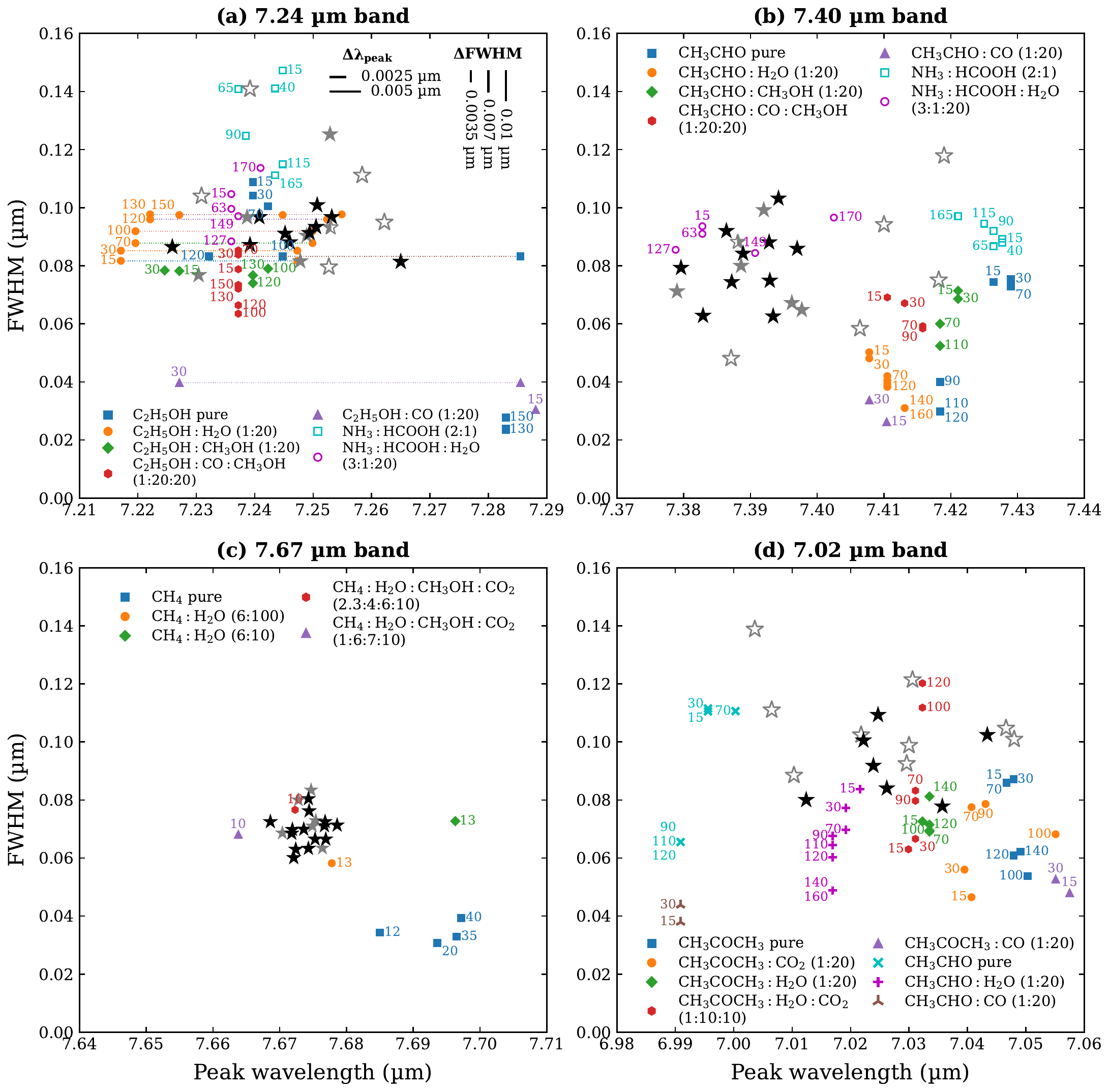}
    \caption{Peak wavelength vs. FWHM of the absorption bands at (a) 7.24~µm, (b) 7.42~µm, (c) 7.67~µm, and (d) 7.02~µm in the JWST/MIRI-MRS spectra and the laboratory spectra of \ce{C2H5OH}, \ce{CH3CHO}, and \ce{HCOO-} ices (labeled as \ce{NH3}:HCOOH mixtures). Values measured from laboratory spectra are denoted by multi-shape markers in different colors (see legend in each panel), and the number next to each marker indicates the temperature in Kelvin. The laboratory data are taken from \cite{TvS2018} for \ce{CH3CHO} and \ce{C2H5OH}, \cite{Rachid2020} for \ce{CH3COCH3}, and this work (Appendix~\ref{appendix:HCOO-_lab_spec}) for \ce{HCOO-}. The legend convention for observational data points is the same as in Fig.~\ref{fig:band_ratios}. In the laboratory spectra of some \ce{C2H5OH} ice mixtures, the band around 7.24~µm is split into two, which are indicated by two data points linked with dotted lines. To avoid clutter, representative uncertainty bars in peak wavelength (0.0025 and 0.005~µm) and FWHM (0.0035, 0.007, and 0.01~µm) are indicated in the upper right corner of panel (a). The observational data points have $\Delta\lambda_\text{peak}=0.005$~µm and $\Delta$FWHM = 0.01~µm; the experimental data points have $\Delta\lambda_\text{peak}=0.0025$~µm and $\Delta$FWHM = 0.0035~µm for \ce{HCOO-} (panels (a), (b)) and \ce{CH3COCH3} ices (panel (d)), and $\Delta\lambda_\text{peak}=0.005$~µm and $\Delta$FWHM = 0.007~µm for \ce{C2H5OH} (panel (a)), \ce{CH3CHO} (panels (b), (d)), and \ce{CH4} ices (panel (c)). See Appendix~\ref{appendix:uncertainty} for the discussion of uncertainty estimation.}
    \label{fig:boogert_plot_7.2-7.7um}
\end{figure*}

In order to better analyze the possible carriers of the observed absorption bands, we measure the peak wavelengths and FWHMs of four bands observed at 7.02, 7.24, 7.40, and 7.67~µm, and compare them with the same quantities derived from laboratory spectra of different ice mixtures (as displayed in Fig.~\ref{fig:boogert_plot_7.2-7.7um}). We first discuss the 7.24, 7.40, and 7.67~µm bands, which have been characterized in previous work (e.g., \citealt{Schutte1999, Yang2022}), and return to the less constrained 7.02~µm band at the end. 

\subsubsection{The 7.24~µm band}\label{sect:results2_7.24}
The 7.24~µm band has scattered peak wavelengths (7.225--7.265~µm) and FWHMs (0.07--0.14~µm) in observations, as shown in Fig.~\ref{fig:boogert_plot_7.2-7.7um}a. The two candidate species contributing to this band are \ce{HCOO^-} and \ce{C2H5OH}. Both pure and \ce{H2O}-mixed \ce{HCOO^-} ices match the observations in terms of peak wavelength, but the FWHM of the \ce{H2O}-mixed one fits better. Here we only plot the more diluted mixture of \ce{HCOO^-} (\ce{NH3}:HCOOH:\ce{H2O} = 3:1:20), given that the values are similar for the other mixture (\ce{NH3}:HCOOH:\ce{H2O} = 2:1:10).

In addition to \ce{HCOO^-}, \ce{C2H5OH} ices at low temperatures, either in pure form or mixed with other common interstellar ices (e.g., \ce{H2O} and \ce{CH3OH}), match the observations well in terms of peak wavelength and FWHM. Only the \ce{CO}-mixed and the crystalline (i.e., high-temperature) \ce{C2H5OH} ices are excluded. This is consistent with the fitting results of the COM ice fingerprint range of B1-c \citep{Chen2024}, which prefer \ce{H2O} and \ce{CH3OH} mixtures of oxygen-bearing COMs, whereas CO mixtures usually have too narrow band widths. However, it should be noted that \ce{C2H5OH} has several other bands with similar intensities in 6.8--7.1~µm and at $\sim$7.5~µm, which would constrain its contribution to the 7.24~µm band in an overall fitting with laboratory spectra.

\subsubsection{The 7.40~µm band}\label{sect:results2_7.40}
The properties of the 7.40~µm band are also scattered in observations (Fig.~\ref{fig:boogert_plot_7.2-7.7um}b). The peak wavelengths vary between 7.38 and 7.425~µm (or 7.40~µm if excluding low-SNR sources), and the FWHMs vary between 0.05 and 0.12~µm (or between 0.06 and 0.11~µm excluding low-SNR sources).
The \ce{HCOO^-}:\ce{H2O} mixture (labeled as HCOOH:\ce{NH3}:\ce{H2O}) fits the observations well in both $\lambda_\text{peak}$ and FWHM, whereas the band of pure \ce{HCOO^-} (labeled as HCOOH:\ce{NH3}) is too redshifted, although with an appropriate FWHM. The good agreement in $\lambda_\text{peak}$ and FWHM at both $\sim$7.24 and $\sim$7.40~µm between observations and the laboratory spectra provides compelling evidence that \ce{HCOO^-} ice is the main carrier of the two observed bands, and it is likely mixed with \ce{H2O} rather than existing in a pure form.

In addition to \ce{HCOO^-}, another candidate carrier of the 7.40~µm band is \ce{CH3CHO}, for which there is an evident difference in $\lambda_\text{peak}$ and FWHM compared with the observations. The \ce{CH3CHO} band is generally narrower and more redshifted; only the \ce{H2O} mixture at low temperatures and the \ce{CO}+\ce{CH3OH} mixture tend to fit the observations. However, even though \ce{CH3CHO} is not likely to be the dominant contributor to the 7.40~µm band, it could still be present in tandem with \ce{HCOO-}, given its potential contribution to another band at 7.02~µm (see Sect.~\ref{sect:results2_7.02}).

\subsubsection{The 7.67~µm (\ce{CH4}) band}\label{sect:results2_7.67}
Different from the 7.24 and 7.40~µm bands, the 7.67~µm band has very consistent $\lambda_\text{peak}$ (mostly 7.67--7.68~µm) and FWHM (mostly 0.06--0.08~µm) in our sample (Fig.~\ref{fig:boogert_plot_7.2-7.7um}c), suggesting a simple contribution from \ce{CH4} ice. Although only a few mixtures of \ce{CH4} have been studied in laboratories, the comparisons in $\lambda_\text{peak}$ and FWHM already show clear evidence that \ce{CH4} ice in protostellar envelopes is not pure but mixed with (or trapped within) other common interstellar molecules such as \ce{H2O}, \ce{CO2}, and \ce{CH3OH}. The mixing conditions are also likely similar from source to source, given the high consistency in observed band properties, although more laboratory data of \ce{CH4} ice mixtures are needed to validate this interpretation. Nevertheless, this shows a good example that systematic analysis of band properties is very helpful for identifying the contributing ice species and their mixing environments.

\subsubsection{The 7.02~µm band}\label{sect:results2_7.02}
The 7.02~µm band is less constrained than the other three bands and shows more scattered peak wavelengths in observations (varying from 7.0 to 7.05~µm; Fig.~\ref{fig:boogert_plot_7.2-7.7um}d). The FWHMs are generally larger (0.08--0.14~µm), consistent with the broad profiles displayed in Fig.~\ref{fig:overview_COM_range}. There are several potential carriers of this band, including \ce{CH3OH}, \ce{CH3CHO}, \ce{C2H5OH}, and \ce{CH3COCH3}. Only \ce{CH3CHO} (the \ce{CH3} deformation mode at $\sim$6.995~µm) and \ce{CH3COCH3} (the \ce{CH3} asymmetric deformation mode at $\sim$7.05~µm) have reported $\lambda_\text{peak}$ and FWHM in the literature \citep{TvS2018, Rachid2020}. This is largely because the \ce{CH3OH} band at $\sim$6.9~µm is wide, multipeaked, and much stronger than the nearby bands of other COMs given its dominant ice abundance, making it very difficult to separate the features of different ices and determine their band properties. 
Nevertheless, the band properties of \ce{CH3OH} mixtures of \ce{CH3CHO}, \ce{C2H5OH}, and \ce{CH3COCH3} ices at $\sim$7.0~µm should not differ too much from those of their CO, \ce{CO2}, or \ce{H2O} mixtures shown in Fig.~\ref{fig:boogert_plot_7.2-7.7um}d. The comparison with observations shows that the \ce{H2O} mixtures of \ce{CH3CHO} and \ce{CH3COCH3} are plausible contributors to this band. Although the FWHM of each of them is slightly smaller than the observations, they can reasonably combine to form a broader band. 

\bigskip
In general, only the 7.67~µm band is contributed by one ice species and shows very consistent peak wavelengths and FWHMs. The other three bands at 7.02, 7.24, and 7.40~µm are more scattered in their properties, which is a reasonable result of being composed of more than one ice band. The favored carriers of each band are consistent with and strengthen the decomposition results in previous case studies (\citealt{Chen2024, Rocha2024, Rayalacheruvu2025, Turner2026}). 
Even without dedicated spectral fitting, this systematic and quantitative analysis of band properties can already provide robust constraints and independent validation for the carriers and compositions underlying the observed band features.


\section{Conclusions}
We present the first study that applies a systematic and quantitative analysis to the four absorption bands in the COM ice fingerprint range of protostellar sources. In particular, we performed comparisons of band properties between high-quality JWST/MIRI-MRS spectra and laboratory spectra of interstellar ice analogs (including newly measured ones for \ce{HCOO-} ices). Such comparisons are one of the most important approaches to identify molecular carriers of observed absorption features and infer their mixing conditions, which can only be achieved with a large sample. We summarize our main findings below:

\begin{enumerate}
    \item The 7.67~µm band is the strongest and most commonly detected absorption band in the COM ice fingerprint range. The two bands at 7.24 and 7.40~µm are well recognized and show up together in most Class 0 and Class 0/I sources, supporting the long-suggested attribution to \ce{HCOO-} ice. 
    
    \item The integrated areas of the three absorption bands at 7.24, 7.40, and 7.67~µm are strongly correlated with each other, suggesting them as common tracers of ice reservoirs in protostellar systems. Nondetection of the 7.24 and 7.40~µm bands is likely due to low ice column densities instead of a true absence of ices in a target.

    \item The integrated area ratios between the 7.24 and the 7.40~µm band suggest additional contributors to the 7.24~µm band beyond \ce{HCOO^-} ice (e.g., \ce{C2H5OH} ice).
    
    \item The 7.67~µm band has remarkably consistent peak wavelengths and FWHMs, suggesting a dominant contribution from \ce{CH4} ice under similar mixing environments. In contrast, the bands at 7.02, 7.24, and 7.40~µm show considerable variation in their band properties, implying contributions from more than one species.
    
    \item By comparing the peak wavelengths and FWHMs of the absorption bands measured from observations and laboratory spectra, we confirmed that \ce{CH4} and \ce{HCOO^-} ices are mixed with other species (e.g., \ce{H2O}) instead of being pure. \ce{HCOO^-} ice mixed with \ce{H2O} is the most likely carrier of the observed 7.24 and 7.40~µm bands. In addition to \ce{HCOO^-}, the 7.24~µm band can also be contributed by \ce{C2H5OH}, although its mixing environment is less constrained. The 7.02 and 7.40~µm bands are likely attributed to \ce{H2O}-mixed \ce{CH3CHO}, with potential contribution from \ce{H2O}-mixed \ce{CH3COCH3} to the 7.02~µm band.
    
    \item The identification of \ce{HCOO-} ice is solidified by both the co-occurrence of the 7.24 and 7.40~µm bands in observations and the agreement in band properties between observations and the new laboratory spectra of \ce{HCOO-}. The presence of \ce{HCOO-} ice as a salt is a strong evidence of ongoing solid-phase chemistry that actively forms complex ices in star-forming regions.

    \item The systematic measurements and comparisons of band properties in this work provide independent constraints on the identification of COM ices (\ce{CH3CHO}, \ce{C2H5OH}, and \ce{CH3COCH3}) without relying on dedicated spectral fitting, while our results remain consistent with those of previous case studies based on spectral fitting.
\end{enumerate}

In summary, this work establishes a framework for interpreting the weak ice features in a quantitative, systematic way, providing more robust constraints on the chemical inventory in the early stages of star and planet formation. We look forward to further elucidating the ubiquity and origin of COMs by applying quantitative analysis of complex ices to more JWST data, and making more gas-to-ice comparisons for the same sources using complementary ALMA data.

\begin{acknowledgments}
This work is based on observations made with the NASA/ESA/CSA James Webb Space Telescope. The data were obtained from the Mikulski Archive for Space Telescopes (MAST) at the Space Telescope Science Institute, which is operated by the Association of Universities for Research in Astronomy, Inc., under NASA contract NAS 5-03127 for JWST. These observations are associated with programs \#1236, \#1290, and \#1802. All the JWST data used in this paper can be found in MAST under DOI:~\dataset[10.17909/mybe-m934]{http://dx.doi.org/10.17909/mybe-m934}. 
Astrochemistry in Leiden is supported by the Netherlands Research School for Astronomy (NOVA), by funding from the European Research Council (ERC) under the European Union's Horizon 2020 research and innovation program ``MOLDISK'' (grant agreement No.~101019751), and by the Danish National Research Foundation through the Center of Excellence ``InterCat'' (grant agreement No.: DNRF150). 
A.C.G. acknowledges support from PRIN-MUR 2022 20228JPA3A ``The path to star and planet formation in the JWST era (PATH)'' funded by NextGeneration EU and by INAF-GoG 2022 ``NIR-dark Accretion Outbursts in Massive Young stellar objects (NAOMY)'' and Large Grant INAF 2022 ``YSOs Outflows, Disks and Accretion: towards a global framework for the evolution of planet forming systems (YODA).'' 
Part of this research by M.N. was carried out at the Jet Propulsion Laboratory, California Institute of Technology, under a contract with the National Aeronautics and Space Administration (80NM0018D0004). 
V.J.M.L.G. acknowledges support by the Spanish program Unidad de Excelencia María de Maeztu CEX2020-001058-M, financed by MCIN/AEI/10.13039/501100011033, and by the MaX-CSIC Excellence Award MaX4-SOMMA-ICE.  
V.J.M.L.G. acknowledges support by the European Research Council (ERC) under the European Union’s Horizon 2020 research and innovation program (grant agreement No. 101098309--PEBBLES).

\facility{JWST.}

\software{\texttt{NumPy} \citep{Numpy2020}, \texttt{Matplotlib} \citep{Matplotlib2007}, \texttt{SciPy} \citep{SciPy2020}, \texttt{LMFIT} \cite{lmfit2014}, \texttt{OpTool} \citep{Dominik2021_optool}, VSCode (\url{https://code.visualstudio.com/}), ChatGPT (\url{https://chat.openai.com/}), GitHub Copilot (\url{https://github.com/features/copilot}).
}
\end{acknowledgments}

\appendix
\section{Method details}\label{appendix:methods}
In this appendix, we provide details and additional information about the two analysis steps and the discarded 7.9--8.8~µm region mentioned in Sect.~\ref{sect:methods}.

\subsection{Removing gas-phase lines}\label{appendix:method_gas_lines}
\begin{figure}[!htb]
    \centering
    \includegraphics[width=\linewidth]{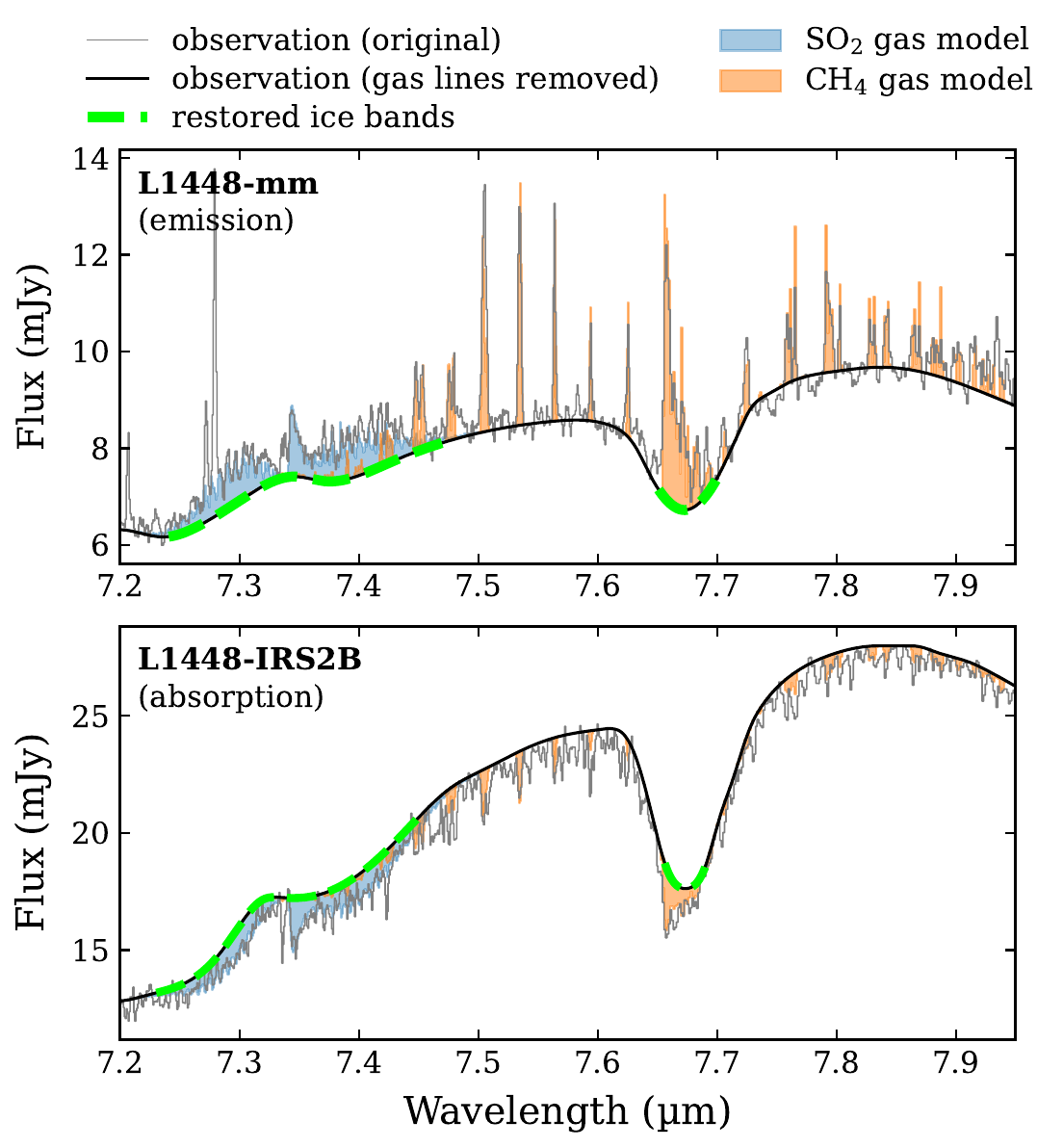}
    \caption{Two examples of how gas-phase lines interfere with the characterization of absorption bands in the COM ice fingerprint range. L1448-mm (top) and L1448-IRS2B (bottom) show the cases of gas-phase lines presented in emission and in absorption, respectively. In each panel, the original observed spectrum and the manually traced baseline of gas-phase lines are plotted in gray and black, respectively. The regions shaded in blue and orange indicate the modeled rovibrational lines of gas-phase \ce{SO2} and \ce{CH4}, respectively. The absorption features contaminated by gas-phase lines are highlighted as dashed green lines.}
    \label{fig:gas_line_example}
\end{figure}

For some sources, the MIRI-MRS spectra extracted at IR continuum peaks are rich in gas-phase lines of simple molecules like \ce{H2O}, \ce{CO2}, and \ce{CH4} (see, e.g., the overview study on gas-phase lines in protostars by \citealt{vanGelder2024_gas}). These lines can be present either in emission or in absorption, superposed on broader ice absorption bands. 
In particular, the P/Q/R-branches of \ce{SO2} gas at $\sim$7.35~µm and the Q-branch of \ce{CH4} gas at 7.66~µm, whether in emission or in absorption, can create a local pseudo-continuum and significantly alter the shape of the absorption bands at 7.24, 7.40, and 7.67~µm (as shown in Fig.~\ref{fig:gas_line_example} for two examples in our sample). A detailed case study on gas-phase \ce{SO2} emission in IRAS~2A by \cite{vanGelder2024_SO2} has also shown the necessity of removing gas lines to restore the absorption features between 7.25 and 7.45~µm. 
Extracting the spectrum at an offset position can be a workaround when studying simple molecules such as \ce{H2O} and \ce{CO2}, which are distributed in more extended regions and have strong ice bands, but this method does not work well for more complex and less abundant species with weak bands. 

In practice, we traced a baseline of gas-phase lines by fitting a univariate spline function to the guiding points selected at line-free channels. This also allows us to smooth the low-SNR part of the spectrum (similar to binning), and clean leftover artifacts that were not fully removed during data reduction \citep[e.g., very strong cosmic-ray showers;][]{Regan2023}.  
In our sample, there are four sources with distinct \ce{SO2} and \ce{CH4} gas lines in emission (L1448-mm, IRAS~1A, IRAS~2A, IRAS~20126), and two in absorption (B1-c and L1448-IRS2B). The rovibrational transitions of \ce{SO2} and \ce{CH4} can be modeled by non-LTE slab models using \texttt{radexpy} (e.g., \citealt{Tabone2023, Francis2024}), and these models are quite sensitive to temperature \citep[e.g.,][]{vanGelder2024_gas}. As a result, it is not difficult to obtain a plausible fit of the rovibrational lines of \ce{SO2} and \ce{CH4} gas by visual inspection. In practice, the fitting was performed species by species (\ce{SO2} followed by \ce{CH4}), and the baseline was traced based on the spectrum with both \ce{SO2} and \ce{CH4} lines subtracted. Figure~\ref{fig:gas_line_example} shows this step for L1448-mm (gas lines in emission) and L1448-IRS2B (gas lines in absorption). We did not model the rovibrational lines of \ce{H2O}, which are abundant but do not show typical P/Q/R-branches that produce pseudo-continuum as symmetric molecules do, thus it is trivial to directly trace a baseline for them by eye.


\begin{figure*}[p]
    \centering
    \begin{subfigure}[b]{0.48\textwidth}
        \centering
        \includegraphics[width=0.98\textwidth]{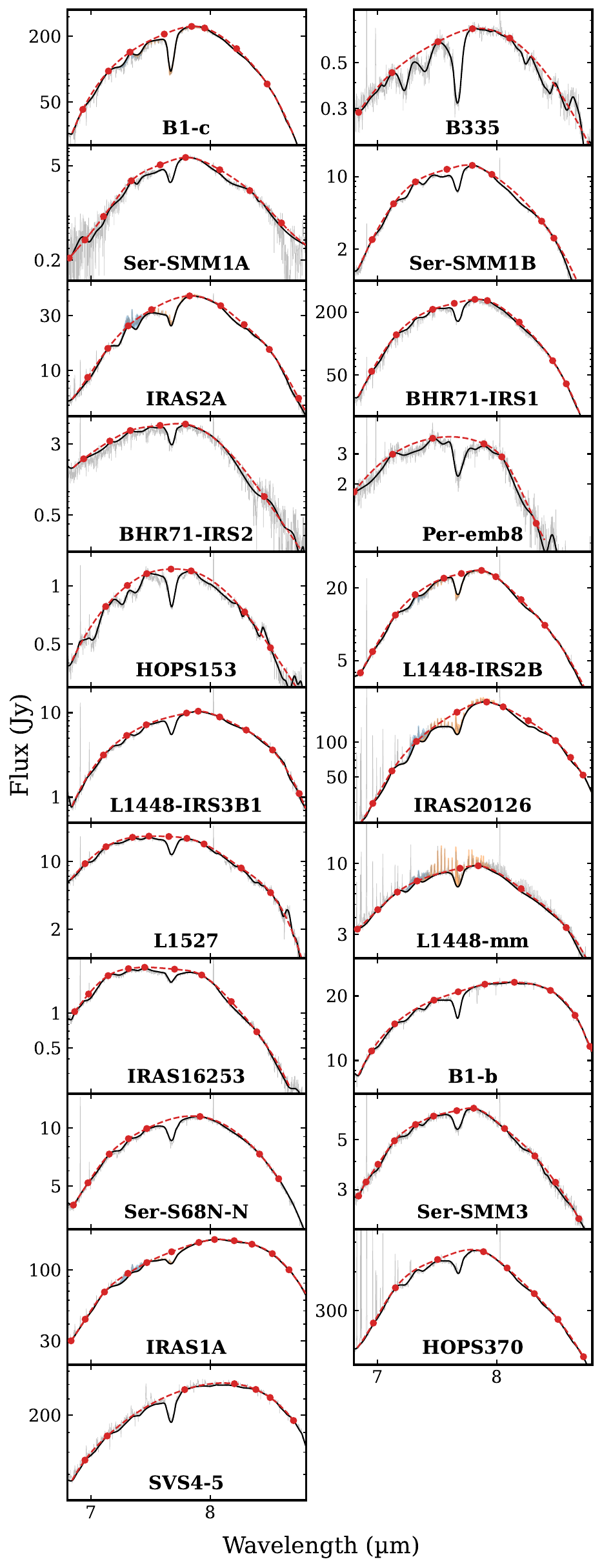}
    \end{subfigure}%
    ~ 
    \begin{subfigure}[b]{0.468\textwidth}
        \centering
        \includegraphics[width=\textwidth]{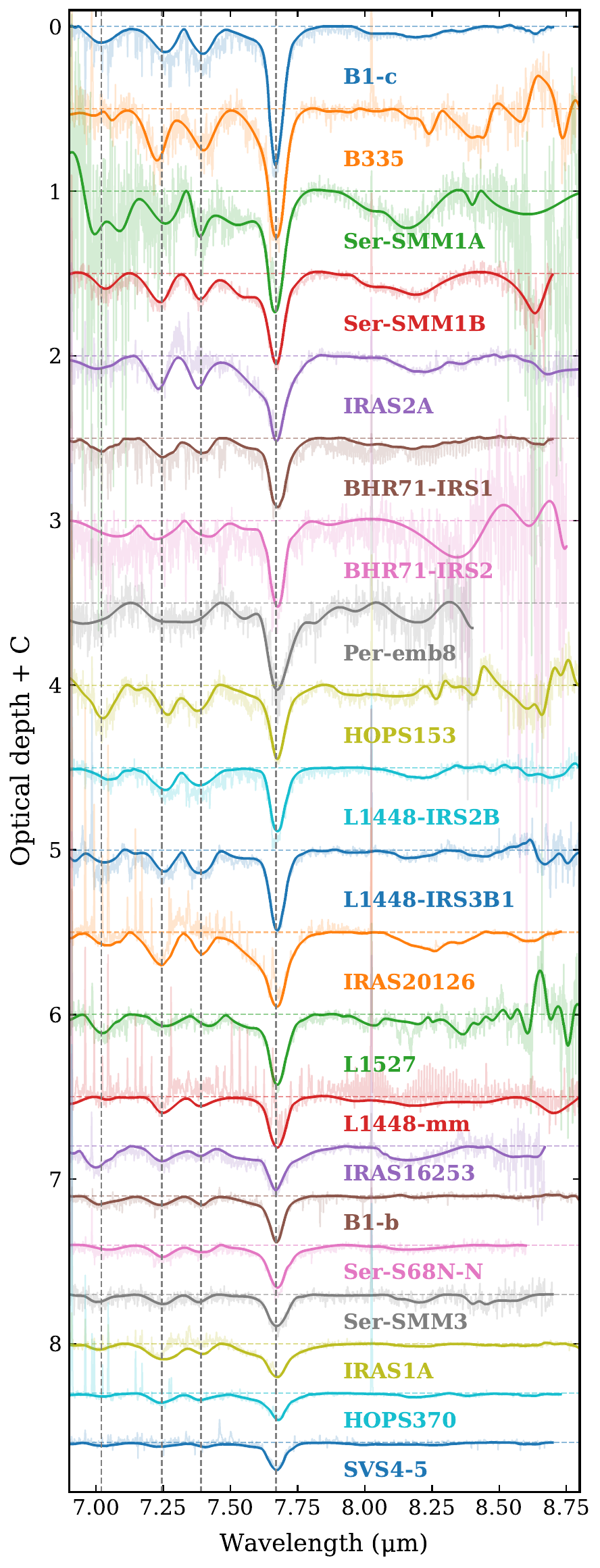}
    \end{subfigure}
    \caption{Left: local continuum traced between 6.8 and 8.8~µm (dashed red lines) and its guiding points (red dots). The observed spectra before and after removing gas-phase lines are plotted in gray and black, respectively. In the panels of six sources (B1-c, IRAS~2A, L1448-IRS2B, IRAS~20126, L1448-mm, IRAS~1A), the modeled rovibrational transitions of gas-phase \ce{SO2} and \ce{CH4} are shaded in blue and orange, respectively (see Sect.~\ref{appendix:method_gas_lines} and Fig.~\ref{fig:gas_line_example} for details). Right: same as Fig.~\ref{fig:overview_COM_range} but for the 6.8--8.8 µm region.}
    \label{fig:full_range}
\end{figure*}


\subsection{Tracing local continua}\label{appendix:method_local_cont}
We traced and subtracted a local continuum for each source to isolate the weak absorption features for further analysis. 
This step was performed on the flux scale, directly after removing the gas-phase lines (Appendix~\ref{appendix:method_gas_lines}). We applied the spline function (same as for removing gas-phase lines), which is more flexible and appropriate than high-order polynomials used in previous studies \citep{Chen2024, Rocha2024}.

The left panel of Fig.~\ref{fig:full_range} shows the selection of the guiding points and the spline fitting of the local continuum for the 21 sources studied in this work. Here we display the full COM ice fingerprint range (i.e., 6.8--8.8~µm) for completeness, although the 7.9--8.8~µm part was not analyzed in this work. We set the guiding points between the major ice bands (e.g., at $\sim$7.15, 7.5, 7.9~µm) at first, and then added a few more points in the middle of the \ce{CH4} band at 7.67~µm to prevent unrealistic curvature, if necessary.  We deliberately set the local continuum as close to the observed spectrum as possible, so that there is almost no absorption left outside the major bands in the isolated spectra. This is slightly different from previous case studies (e.g., \citealt{Chen2024, Rocha2024, Gieser2026}), where there is usually some weak absorption attributed to O-COMs in the inter-band regions (e.g., around 7.15, 7.35, 7.50~µm). However, without a full decomposition of the 6.8--8.8~µm region, we are unable to know how deep these weak features are, or in other words, how much we should lift the local continua from the observed spectra of different sources. In order to keep the band property measurements as consistent as possible across the whole sample, we took the observed spectra as benchmarks when setting the guiding points.

We are aware of the possibility of underestimating the optical depths of the isolated absorption features by doing this, and the measured $\tau_\text{peak}$ and $I_\text{band}$ should be regarded as lower limits. However, some of the band properties, such as $\lambda_\text{peak}$ and $I_\text{band}$ ratios, are less sensitive to the choice of local continua. Taking B1-c as an example, we compared the band properties measured in this work and \cite{Chen2024} using the same method (i.e., Gaussian fitting). Since the local continuum was placed at a more lifted position in \cite{Chen2024}, the corresponding FWHM, $\tau_\text{peak}$, and $I_\text{band}$ are expected to be larger. We found that the $\lambda_\text{peak}$ of the four absorption bands remains almost the same, with a difference of $\lesssim0.005$~µm or 0.1\%. The difference in FWHM is also $\lesssim$0.01~µm or 10\%, except for the 7.02~µm band ($\sim$25\%). This amount of difference in $\lambda_\text{peak}$ and FWHM in observational data is generally within the estimated error bars (Appendix~\ref{appendix:uncertainty_obs}), and will not change our interpretation of Fig.~\ref{fig:boogert_plot_7.2-7.7um}. 


\subsection{The 7.9--8.8 µm region}\label{appendix:7.9-8.8um}
We mentioned in Sect.~\ref{sect:methods} that the 7.9--8.8~µm was not analyzed in this work because the absorption features in this wavelength range are relatively weak and irregular. Previous case studies \citealt{Chen2024, Rocha2024, Gieser2026} suggest several carriers for these features, including \ce{H2CO} (8.02~µm), HCOOH (8.23~µm), \ce{CH3OCHO} (8.26~µm), \ce{CH3OCH3} (8.63~µm), and potentially \ce{CH3COCH3} (8.03~µm). 
However, as shown in both panels of Fig.~\ref{fig:full_range}, only a few sources display clear absorption features in this range (e.g., B1-c, Ser-SMM1B, IRAS\,2A, BHR\,71-IRS\,1, L1448-IRS2B, and IRAS\,20126). 
Other sources are either not ice-rich enough to exhibit detectable COM ice absorption features because of the low-mass nature or late evolutionary stages, or have insufficient SNRs due to the deep silicate band at 9.8~µm and/or limited observing time. Even for those high-SNR sources, their absorption features are multipeaked (i.e., likely a blend of multiple components), making the observed band properties less representative in ice identification. As a result, it is difficult to extract useful statistics and perform subsequent analyses for this wavelength region.

Although the 7.9--8.8~$\mu$m range was excluded from the analysis in this work, we do not expect its inclusion to qualitatively change our results. Firstly, the local continuum placement in the 6.8--7.9~µm region is largely independent of the excluded 7.9--8.8~µm region. The two regions are separated by the deep \ce{CH4} band, while the absorption around 7.9~µm is minimal, making the continuum anchor at this wavelength relatively robust. Secondly, we did not claim any definitive identification for the ice species relevant to this range (e.g., \ce{CH3COCH3}. While extending the analysis to the 7.9--8.8 µm region could potentially provide additional supporting or opposing evidence, the exclusion of this wavelength range does not affect our current conclusions. We hope to extend this systematic analysis to the 7.9--8.8 µm region using samples with more robust detections in the future.

\section{Table of band properties in the observations}\label{appendix:table_band_prop}
Table~\ref{tab:band_prop} lists the properties of the four observed absorption bands at 7.02, 7.24, 7.40, and 7.67~µm. There are six columns for each band: the peak wavelength ($\lambda_\text{peak}$), the FWHM, the peak optical depth ($\tau_\text{peak}$), the noise level in optical depth ($\tau_\sigma$), the integrated band area, and the fitting method (by Gaussian or by definition).

\begin{sidewaystable*}[!hp]
    \setlength{\tabcolsep}{0.11cm}
    \renewcommand{\arraystretch}{1.1}
    \centering
    \caption{Band Properties of the Observed Absorption Bands at 7.02, 7.24, 7.40, and 7.67~µm.}
    \begin{tabular}{l|c@{\hspace{3pt}}ccccc|c@{\hspace{3pt}}ccccc|c@{\hspace{3pt}}ccccc|c@{\hspace{3pt}}ccccc}
    \hline\hline
    \multirow{3}{*}{\small \bf Source} & \multicolumn{6}{c|}{\bf 7.02~µm band} & \multicolumn{6}{c|}{\bf 7.24~µm band} & \multicolumn{6}{c|}{\bf 7.40~µm band} & \multicolumn{6}{c}{\bf 7.67~µm band}\\
    & $\lambda_\text{peak}$ & {\scriptsize FWHM} & $\tau_\text{peak}$ & $\tau_\sigma$$^\text{a}$ & $I_\text{band}$ & Fit$^\text{b}$
    & $\lambda_\text{peak}$ & {\scriptsize FWHM} & $\tau_\text{peak}$ & $\tau_\sigma$$^\text{a}$ & $I_\text{band}$ & Fit$^\text{b}$
    & $\lambda_\text{peak}$ & {\scriptsize FWHM} & $\tau_\text{peak}$ & $\tau_\sigma$$^\text{a}$ & $I_\text{band}$ & Fit$^\text{b}$ 
    & $\lambda_\text{peak}$ & {\scriptsize FWHM} & $\tau_\text{peak}$ & $\tau_\sigma$$^\text{a}$ & $I_\text{band}$ & Fit$^\text{b}$ \\
    & (µm) & (µm) & & & {\scriptsize (cm$^{-1}$)} & & 
    (µm) & (µm) & & & {\scriptsize (cm$^{-1}$)} & & 
    (µm) & (µm) & & & {\scriptsize (cm$^{-1}$)} & & 
    (µm) & (µm) & & & {\scriptsize (cm$^{-1}$)} & \\
    \hline
    B1-b & 7.0306 & 0.1214 & 0.05 & 1.1(-2) & 1.29 & G & 7.2408 & 0.0968 & 0.06 & 8.3(-3) & 1.12 & G & 7.3934 & 0.0626 & 0.06 & 8.1(-3) & 0.68 & G & 7.6724 & 0.0630 & 0.26 & 7.3(-3) & 3.02 & G \\ 
    B1-c & 7.0247 & 0.1093 & 0.10 & 3.1(-3) & 2.22 & G & 7.2532 & 0.0968 & 0.16 & 3.1(-3) & 3.08 & G & 7.3970 & 0.0859 & 0.17 & 3.1(-3) & 2.80 & G & 7.6706 & 0.0584 & 0.80 & 3.1(-3) & 8.44 & G \\ 
    IRAS 1A & 7.0124 & 0.0800 & 0.04 & 5.4(-3) & 0.61 & G & 7.2529 & 0.1253 & 0.10 & 5.4(-3) & 2.46 & G & 7.3977 & 0.0648 & 0.06 & 5.4(-3) & 0.77 & G & 7.6728 & 0.0800 & 0.18 & 2.8(-3) & 2.67 & G \\ 
    IRAS 2A & 7.0036 & 0.1388 & 0.08 & 1.7(-2) & 2.36 & G & 7.2304 & 0.0768 & 0.20 & 1.7(-2) & 3.10 & G & 7.3790 & 0.0712 & 0.20 & 1.7(-2) & 2.80 & D & 7.6754 & 0.0729 & 0.46 & 8.1(-3) & 6.03 & G \\ 
    L1448-mm & -- & -- & -- & -- & -- & -- & 7.2478 & 0.0816 & 0.10 & 1.1(-2) & 1.55 & D & 7.3886 & 0.0800 & 0.06 & 1.1(-2) & 0.89 & D & 7.6749 & 0.0710 & 0.31 & 1.1(-2) & 4.01 & G \\ 
    L1448-IRS2B & 7.0466 & 0.1047 & 0.07 & 6.9(-3) & 1.62 & G & 7.2489 & 0.0904 & 0.14 & 5.7(-3) & 2.52 & G & 7.3920 & 0.0992 & 0.11 & 5.7(-3) & 2.15 & G & 7.6764 & 0.0634 & 0.39 & 4.4(-3) & 4.51 & G \\ 
    L1448-IRS3B1 & 7.0296 & 0.0925 & 0.08 & 1.6(-2) & 1.48 & G & 7.2527 & 0.0796 & 0.13 & 1.2(-2) & 2.02 & G & 7.3881 & 0.0882 & 0.15 & 1.2(-2) & 2.43 & G & 7.6753 & 0.0665 & 0.48 & 7.5(-3) & 5.79 & G \\ 
    Per-emb 8 & -- & -- & -- & -- & -- & -- & -- & -- & -- & -- & -- & -- & -- & -- & -- & -- & -- & -- & 7.6779 & 0.0875 & 0.53 & 7.2(-2) & 8.34 & G \\ 
    \hline
    L1527 & 7.0262 & 0.0840 & 0.11 & 1.4(-2) & 2.07 & G & 7.2622 & 0.0950 & 0.06 & 1.4(-2) & 1.10 & G & 7.4182 & 0.0751 & 0.07 & 1.4(-2) & 0.94 & G & 7.6768 & 0.0725 & 0.42 & 1.2(-2) & 5.53 & G \\ 
    \hline
    Ser-S68N-N & 7.0480 & 0.1009 & 0.03 & 1.0(-2) & 0.62 & G & 7.2494 & 0.0913 & 0.07 & 9.4(-3) & 1.27 & G & 7.3928 & 0.0881 & 0.05 & 9.4(-3) & 0.69 & G & 7.6744 & 0.0762 & 0.24 & 9.7(-3) & 3.27 & G \\ 
    Ser-SMM1A & -- & -- & -- & -- & -- & -- & 7.2584 & 0.1112 & 0.20 & 8.5(-2) & 4.06 & D & 7.3871 & 0.0481 & 0.28 & 8.5(-2) & 2.66 & G & 7.6704 & 0.0686 & 0.67 & 3.2(-2) & 8.26 & G \\ 
    Ser-SMM1B & 7.0357 & 0.0778 & 0.09 & 8.5(-3) & 1.57 & G & 7.2392 & 0.0872 & 0.17 & 7.2(-3) & 2.95 & D & 7.3872 & 0.0744 & 0.16 & 7.2(-3) & 2.14 & D & 7.6718 & 0.0685 & 0.51 & 2.9(-3) & 6.37 & G \\ 
    Ser-SMM3 & 7.0103 & 0.0885 & 0.04 & 3.2(-2) & 0.85 & G & 7.2460 & 0.0881 & 0.06 & 2.5(-2) & 1.03 & G & 7.3829 & 0.0628 & 0.05 & 1.7(-2) & 0.55 & G & 7.6743 & 0.0804 & 0.19 & 1.6(-2) & 2.83 & G \\ 
    SVS4-5 & 7.0300 & 0.0988 & 0.02 & 1.5(-3) & 0.43 & G & 7.2392 & 0.1409 & 0.03 & 1.5(-3) & 0.72 & G & 7.4064 & 0.0584 & 0.02 & 1.5(-3) & 0.27 & G & 7.6719 & 0.0698 & 0.16 & 4.8(-3) & 2.08 & G \\ 
    \hline
    HOPS 153 & 7.0239 & 0.0918 & 0.20 & 2.8(-2) & 3.92 & G & 7.2650 & 0.0814 & 0.17 & 2.8(-2) & 2.87 & G & 7.3796 & 0.0793 & 0.16 & 2.6(-2) & 2.51 & G & 7.6769 & 0.0665 & 0.42 & 9.4(-3) & 5.06 & G \\ 
    HOPS 370 & 7.0218 & 0.1024 & 0.02 & 4.8(-4) & 0.42 & G & 7.2452 & 0.0911 & 0.06 & 4.8(-4) & 1.09 & G & 7.3864 & 0.0920 & 0.04 & 4.8(-4) & 0.73 & D & 7.6786 & 0.0713 & 0.15 & 1.1(-3) & 1.90 & G \\ 
    \hline
    BHR71-IRS1 & 7.0222 & 0.1005 & 0.08 & 3.4(-3) & 1.66 & G & 7.2505 & 0.0934 & 0.12 & 3.4(-3) & 2.18 & G & 7.3929 & 0.0749 & 0.09 & 3.4(-3) & 1.37 & G & 7.6767 & 0.0711 & 0.42 & 2.6(-3) & 5.44 & G \\ 
    BHR71-IRS2 & -- & -- & -- & -- & -- & -- & 7.2309 & 0.1040 & 0.12 & 5.0(-2) & 2.48 & G & 7.4100 & 0.0942 & 0.10 & 5.0(-2) & 1.89 & G & 7.6743 & 0.0633 & 0.51 & 3.4(-2) & 5.79 & G \\ 
    \hline
    B335 & -- & -- & -- & -- & -- & -- & 7.2259 & 0.0865 & 0.31 & 3.9(-2) & 5.41 & G & 7.3942 & 0.1032 & 0.25 & 3.9(-2) & 4.95 & G & 7.6736 & 0.0699 & 0.78 & 1.5(-2) & 9.82 & G \\ 
    IRAS 16253 & 7.0065 & 0.1110 & 0.13 & 1.5(-2) & 3.02 & G & 7.2507 & 0.1009 & 0.09 & 1.0(-2) & 1.85 & G & 7.3889 & 0.0842 & 0.06 & 1.0(-2) & 0.91 & G & 7.6686 & 0.0725 & 0.22 & 3.5(-3) & 2.93 & G \\ 
    IRAS 20126 & 7.0434 & 0.1024 & 0.08 & 4.6(-3) & 1.74 & G & 7.2387 & 0.0967 & 0.20 & 5.3(-3) & 3.94 & G & 7.3962 & 0.0672 & 0.14 & 5.3(-3) & 1.78 & G & 7.6747 & 0.0834 & 0.42 & 1.9(-3) & 6.36 & G \\ 
    \hline
    \end{tabular}
    \begin{minipage}{\textwidth}
        (1) Because the local continua are conservatively placed near the observed spectra (see Appendix~\ref{appendix:method_local_cont}), $\tau_\text{peak}$ and area listed here should be considered as lower limits.\\
        (2) The uncertainties of $\lambda_\text{peak}$, FWHM, and area are estimated as 0.005~µm, 0.01~µm, and 20\%, respectively (see Appendix~\ref{appendix:uncertainty_obs} for discussion).\\
        $^\text{a}$ $\tau_\sigma$ are shown in the form $a$($b$), which denotes $a\times10^\text{b}$.\\
        $^\text{b}$ The ``fit'' column lists how the band was fitted and the properties were derived: G = by Gaussian fitting, D = by definition of peak and FWHM.      
    \end{minipage}
    \label{tab:band_prop}
\end{sidewaystable*}

\section{Uncertainty estimation of band properties}\label{appendix:uncertainty}
\subsection{Observed spectra}\label{appendix:uncertainty_obs}
The uncertainties in the Gaussian-fitted $\lambda_\text{peak}$ and FWHM (Sect.~\ref{sect:methods}) are relatively small, only on the order of $10^{-4}$~µm. For sources with decent SNRs, the peak wavelength uncertainty ($\Delta\lambda_\text{peak}$) is about (1--2)$\times10^{-4}$~µm, and the FWHM uncertainty ($\Delta$FWHM) is about (3--5)$\times10^{-4}$~µm. For those with low SNRs, $\Delta\lambda_\text{peak}$ and $\Delta$FWHM can reach up to $\sim$5$\times10^{-4}$ and $\sim$1$\times10^{-3}$~µm, respectively. However, larger uncertainties should have been introduced by removing the gas-phase lines and subtracting the local continuum (Appendices~\ref{appendix:method_gas_lines}--\ref{appendix:method_local_cont}), which cannot be quantified analytically. Although the actual uncertainties differ from source to source, we adopt a uniform and conservative estimate for $\Delta\lambda_\text{peak}$ and $\Delta$FWHM, which are 0.005~µm and 0.01~µm (i.e., one order of magnitude larger than the typical uncertainties of the Gaussian fitting), respectively. The estimated $\Delta\lambda_\text{peak}$ and $\Delta$FWHM for observed bands are denoted as black bars in Fig.~\ref{fig:boogert_plot_7.2-7.7um}a; they are also shown as shaded regions in Fig.~\ref{fig:72-74um_fit}, which look reasonable. 

The uncertainties of the band area are even more complicated to estimate. Assuming a Gaussian profile, the band area can be calculated analytically from 
\begin{equation}\label{eq:band_area}
    I_\text{band} = \int_{\tilde{\nu_1}}^{\tilde{\nu_2}}  \tau(\tilde{\nu})\,\mathrm{d}\tilde{\nu} = \sqrt{2\pi}\,\tau_\text{peak}\sigma = \sqrt{\frac{\pi}{4\ln2}}\,\tau_\text{peak}\,\text{FWHM},
\end{equation}
where $\tilde{\nu}$ is the wavenumber, $\tau_\text{peak}$ is the peak optical depth, and $\sigma=\text{FWHM}/2\sqrt{\ln2}$. The error propagation then gives 
\begin{equation}\label{eq:band_area_err}
    \frac{\Delta I_\text{band}}{I_\text{band}} =  \sqrt{\left(\frac{\Delta\tau_\text{peak}}{\tau_\text{peak}}\right)^2+\left(\frac{\Delta\text{FWHM}}{\text{FWHM}}\right)^2}.
\end{equation}
The relative uncertainty of FWHM, when considering the conservative estimate (0.01~µm), is around 10\%--20\%. The relative uncertainty of $\Delta\tau_\text{peak}$  obtained from Gaussian fitting itself is small (well below 1\%). If we estimate $\Delta\tau_\text{peak}/\tau_\text{peak}$ to be 10\%, the relative uncertainty of band area would be around 20\%, which is adopted in Fig.~\ref{fig:band_ratios} for data points in observations.

\subsection{Laboratory spectra}\label{appendix:uncertainty_lab}
We assume that the uncertainties in $\lambda_\text{peak}$ and FWHM of laboratory spectra are dominated by measurement uncertainties (i.e., the spectral resolution of experiments). Conventionally, the laboratory spectra are measured in wavenumber, so we need to convert the uncertainties from wavenumber to wavelength via $\Delta\lambda=\lambda^2\Delta\tilde{\nu}$. For \ce{CH3CHO} and \ce{C2H5OH} ices, the spectral resolution is $\Delta\tilde{\nu}=1$~cm$^{-1}$ \citep{TvS2018}, which corresponds to $\Delta\lambda\sim0.005$--0.006~µm in 7--8~µm. The spectral resolution for \ce{HCOO-} is better (Appendix~\ref{appendix:HCOO-_experiment}), with $\Delta\tilde{\nu}=0.5$~cm$^{-1}$ and the corresponding $\Delta\lambda\sim0.0025$--0.003~µm. The absorption band widths of the laboratory spectra are measured based on the definition of FWHM (i.e., $\tilde{\nu}_2-\tilde{\nu}_1$) instead of fitting a Gaussian function, so the propagated uncertainty is given by $\Delta$FWHM$_{\tilde{\nu}} = \sqrt{\Delta\tilde{\nu}^2_1+\Delta\tilde{\nu}^2_2} = \sqrt{2}\Delta\tilde{\nu}$. The factor of $\sqrt{2}$ (or $1.41$) is the same between $\Delta$FWHM$_\lambda$ and $\Delta\lambda$. As a result, the FWHM uncertainty is 0.007--0.0084~µm for the \ce{CH3CHO} and \ce{C2H5OH} spectra, and 0.0035--0.0042~µm for the \ce{HCOO-} ones. The uncertainty bars for laboratory spectra in Fig.~\ref{fig:boogert_plot_7.2-7.7um}a correspond to 0.003 and 0.006~µm for $\Delta\lambda$, and 0.0084 and 0.0042~µm for $\Delta$FWHM.

According to Eq.~\eqref{eq:band_area_err}, the relative uncertainties of band area ($I_\text{band}$) depend on the relative uncertainties in peak optical depth ($\tau_\text{peak}$) and FWHM, of which the latter is generally smaller than 10\%. Usually, $\tau_\text{peak}$ is not measured and reported by experimentalists, as it is proportional to the column densities of the deposited ices, but its relative uncertainty should be fairly small (e.g., smaller than observational uncertainties) given the high SNRs of laboratory spectra. The main uncertainty of $\tau_\text{peak}$ should come from the baseline correction (see Appendix~\ref{appendix:HCOO-_band_prop}), but would still not exceed 10\%--20\%. As a result, a relative uncertainty of 15\% should be a reasonable conservative estimate for an $I_\text{band}$ measured from laboratory spectra, and is adopted in the left panel of Fig.~\ref{fig:band_ratios}.

\section{Laboratory spectra of \ce{HCOO^-} ices}\label{appendix:HCOO-_lab_spec}
\subsection{Experiments}\label{appendix:HCOO-_experiment}
New laboratory IR spectra of ice mixtures of NH$_{3}$:HCOOH (2:1) and \ce{NH3}:HCOOH:\ce{H2O} (2:1:10 and 3:1:20) were collected using the InfraRed Absorption Setup for Ice Spectroscopy (IRASIS) in the Leiden Laboratory for Astrophysics. The setup, calibration, and deposition methods are described in \cite{Slavicinska2025_NH4SH} and references therein. Briefly, three dosing lines with independent leak valves were calibrated individually to deposit either pure \ce{H2O}, \ce{NH3}, or HCOOH on a KBr substrate held at 15~K. Ice deposition rates for each species were chosen so that, once all three leak valves were opened simultaneously, the deposited ice mixture would have the desired mixing ratio. Deposition rates during calibrations (described in Appendix A in \citealt{Slavicinska2025_NH4SH}) were determined via laser interferometry, using the ice refractive indexes and densities presented in Table~\ref{tab:refractive_index}.

\begin{table}[!h]
    \setlength{\tabcolsep}{0.07cm}
    \centering
    \caption{Ice Refractive Indexes and Densities Used in the Calibration.}
    \begin{tabular}{cccc}
    \hline\hline
    Molecules & $n$ & $\rho$ (g cm$^{-3}$) & References \\
    \hline
    \ce{H2O}	& 1.234 & 0.719	&		\cite{Yarnall2022}\\
    \ce{NH3} & 1.33 & 0.68 & \cite{Hudson2022} \\
    \ce{HCOOH} & 1.291 & 0.979 & \cite{Hudson2022}\\
    \hline
    \end{tabular}
    \label{tab:refractive_index}
\end{table}

HCOOH was chosen as the limiting reactant in the acid-base reaction \ce{NH3} + HCOOH $\rightarrow$ \ce{NH4HCOO} to minimize the contribution of neutral HCOOH to the CH bending mode at 7.24~µm, a shared feature between HCOOH and \ce{HCOO-}. The formation of some \ce{NH4HCOO} was already observed during all of the ice mixture depositions at 15 K, although strong features at 5.85~µm (HCOOH C=O stretch) and 8.22~µm (HCOOH C--O stretch) indicated that the reaction remained far from complete at this temperature. After deposition, the ice mixture was heated at a rate of 1 K min$^{-1}$, during which the peaks characteristic of \ce{NH4HCOO} grew while the peaks of the neutral reactants diminished. IR spectra (0.5 cm$^{-1}$ resolution, 256 scans per spectrum) of the ice mixture were continuously collected throughout heating, resulting in an average ice temperature difference of $\sim$3.6 K between each spectrum.

\begin{figure*}[!htp]
    \centering
    \includegraphics[width=0.9\textwidth]{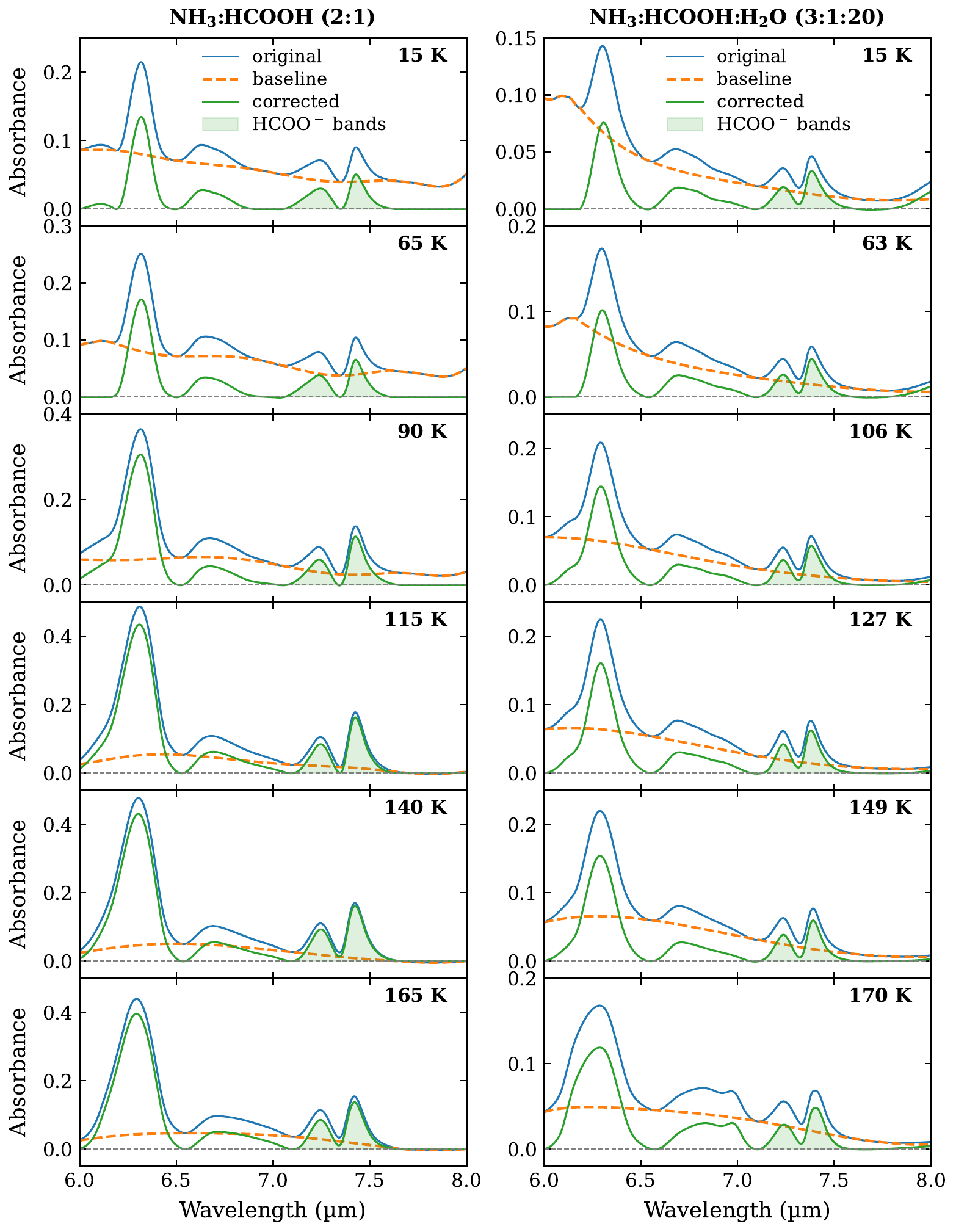}
    \caption{Baseline correction of the laboratory IR spectra of two ice mixtures, \ce{NH3}:HCOOH = 2:1 (left column) and \ce{NH3}:HCOOH:\ce{H2O} = 3:1:20 (right column). In each panel, the original and baseline-corrected spectra are plotted in blue and green, respectively. The polynomial baseline is shown as the dashed orange line. The two \ce{HCOO-} bands of interest are shaded in green. The measurement temperature is indicated in the upper right corner.}
    \label{fig:HCOO-_blcorr}
\end{figure*}

\subsection{Baseline correction and band property measurements}\label{appendix:HCOO-_band_prop}
The laboratory IR spectra of the aforementioned ice mixtures were baseline corrected by fitting a fifth- to seventh-order polynomial to manually selected wavelength ranges free of absorption features. Figure~\ref{fig:HCOO-_blcorr} shows the original and the baseline-corrected spectra of \ce{NH3}:HCOOH (2:1) and \ce{NH3}:HCOOH:\ce{H2O} (3:1:20). The baseline correction was applied only to the $\sim$6.0--8.0~µm region, where the two  \ce{HCOO-} bands at 7.24 and 7.39~µm are located. The peak wavelength, FWHM, and integrated area of the two \ce{HCOO-} bands were measured from the baseline-corrected spectra of three ice mixtures at six different temperatures (see Table~\ref{tab:HCOO-_band_prop}).

\begin{table}[!htb]
    \centering
    \caption{Properties of the Two \ce{HCOO-} Bands at $\sim$7.24~µm (Columns (2)--(4)) and $\sim$7.39~µm (Columns (5)--(7)) in Three Ice Mixtures.}
    \begin{tabular}{c|ccc|ccc}
    \hline\hline
    $T$ & Peak $\lambda$ & FWHM & $I_\text{band}$ & Peak $\lambda$ & FWHM & $I_\text{band}$ \\
    (K) & (µm) & (µm) & (cm$^{-1}$) & (µm) & (µm) & (cm$^{-1}$) \\ \hline
    \multicolumn{7}{c}{\bf \ce{NH3}:HCOOH (2:1)}\\ \hline
    15 & 7.2448 & 0.1473 & 1.9574 & 7.4277 & 0.0893 & 2.0712 \\ 
    65 & 7.2372 & 0.1409 & 2.4090 & 7.4264 & 0.0867 & 2.6332 \\ 
    90 & 7.2385 & 0.1260 & 3.3473 & 7.4264 & 0.0920 & 4.9887 \\ 
    115 & 7.2448 & 0.1150 & 4.3326 & 7.4250 & 0.0945 & 7.0366 \\ 
    140 & 7.2473 & 0.1164 & 4.8702 & 7.4237 & 0.0998 & 7.5005 \\ 
    165 & 7.2435 & 0.1112 & 3.9304 & 7.4211 & 0.0971 & 5.5752 \\ \hline 
    \multicolumn{7}{c}{\bf \ce{NH3}:HCOOH:\ce{H2O} (2:1:10)} \\ \hline
    15 & 7.2360 & 0.1058 & 1.6257 & 7.3867 & 0.0950 & 2.7346 \\ 
    55 & 7.2347 & 0.1034 & 1.9995 & 7.3854 & 0.0937 & 3.3921 \\ 
    98 & 7.2360 & 0.0984 & 2.7578 & 7.3841 & 0.0936 & 4.4609 \\ 
    120 & 7.2372 & 0.0972 & 3.3015 & 7.3841 & 0.0923 & 5.0103 \\ 
    141 & 7.2372 & 0.0959 & 3.5890 & 7.3867 & 0.0910 & 5.1122 \\ 
    163 & 7.2435 & 0.1176 & 3.5441 & 7.3999 & 0.0939 & 4.7266 \\ \hline 
    \multicolumn{7}{c}{\bf \ce{NH3}:HCOOH:\ce{H2O} (3:1:20)} \\ \hline
    15 & 7.2360 & 0.1047 & 0.9131 & 7.3828 & 0.0936 & 1.4999 \\ 
    63 & 7.2360 & 0.0996 & 1.1835 & 7.3828 & 0.0910 & 1.9232 \\ 
    106 & 7.2360 & 0.0921 & 1.5502 & 7.3814 & 0.0909 & 2.4938 \\ 
    127 & 7.2360 & 0.0884 & 1.7104 & 7.3788 & 0.0855 & 2.6040 \\ 
    149 & 7.2372 & 0.0971 & 1.7162 & 7.3907 & 0.0844 & 2.4088 \\ 
    170 & 7.2410 & 0.1137 & 1.4682 & 7.4025 & 0.0966 & 2.1323 \\ 
    \hline 
    \end{tabular}
    \label{tab:HCOO-_band_prop}
\end{table}

\clearpage
\bibliography{references}{}

@ARTICLE{Herbst2009,
       author = {{Herbst}, Eric and {van Dishoeck}, Ewine F.},
        title = "{Complex Organic Interstellar Molecules}",
      journal = {\araa},
         year = 2009,
        month = sep,
       volume = {47},
       number = {1},
        pages = {427-480},
          doi = {10.1146/annurev-astro-082708-101654},
       adsurl = {https://ui.adsabs.harvard.edu/abs/2009ARA&A..47..427H}
}

@INPROCEEDINGS{Ceccarelli2023,
       author = {{Ceccarelli}, C. and {Codella}, C. and {Balucani}, N. and {Bockelee-Morvan}, D. and {Herbst}, E. and {Vastel}, C. and {Caselli}, P. and {Favre}, C. and {Lefloch}, B. and {Oberg}, K. and {Yamamoto}, S.},
        title = "{Organic Chemistry in the First Phases of Solar-Type Protostars}",
    booktitle = {Protostars and Planets VII},
         year = 2023,
       editor = {{Inutsuka}, S. and {Aikawa}, Y. and {Muto}, T. and {Tomida}, K. and {Tamura}, M.},
       series = {Astronomical Society of the Pacific Conference Series},
       volume = {534},
        month = jul,
        pages = {379},
       adsurl = {https://ui.adsabs.harvard.edu/abs/2023ASPC..534..379C}
}

@ARTICLE{Jorgensen2020,
       author = {{J{\o}rgensen}, Jes K. and {Belloche}, Arnaud and {Garrod}, Robin T.},
        title = "{Astrochemistry During the Formation of Stars}",
      journal = {\araa},
         year = 2020,
        month = aug,
       volume = {58},
        pages = {727-778},
          doi = {10.1146/annurev-astro-032620-021927},
archivePrefix = {arXiv},
       eprint = {2006.07071},
 primaryClass = {astro-ph.SR},
       adsurl = {https://ui.adsabs.harvard.edu/abs/2020ARA&A..58..727J}
}

@ARTICLE{Balucani2015,
       author = {{Balucani}, N. and {Ceccarelli}, C. and {Taquet}, V.},
        title = "{Formation of complex organic molecules in cold objects: the role of gas-phase reactions.}",
      journal = {\mnras},
         year = 2015,
        month = apr,
       volume = {449},
        pages = {L16-L20},
          doi = {10.1093/mnrasl/slv009},
archivePrefix = {arXiv},
       eprint = {1501.03668},
 primaryClass = {astro-ph.SR},
       adsurl = {https://ui.adsabs.harvard.edu/abs/2015MNRAS.449L..16B}
}

@ARTICLE{Garrod2022,
       author = {{Garrod}, Robin T. and {Jin}, Miwha and {Matis}, Kayla A. and {Jones}, Dylan and {Willis}, Eric R. and {Herbst}, Eric},
        title = "{Formation of Complex Organic Molecules in Hot Molecular Cores through Nondiffusive Grain-surface and Ice-mantle Chemistry}",
      journal = {\apjs},
         year = 2022,
        month = mar,
       volume = {259},
       number = {1},
          eid = {1},
        pages = {1},
          doi = {10.3847/1538-4365/ac3131},
archivePrefix = {arXiv},
       eprint = {2110.09743},
 primaryClass = {astro-ph.GA},
       adsurl = {https://ui.adsabs.harvard.edu/abs/2022ApJS..259....1G}
}

@ARTICLE{vD2025_JOYS,
       author = {{van Dishoeck}, E.~F. and {Tychoniec}, {\L}. and {Rocha}, W.~R.~M. and {Slavicinska}, K. and {Francis}, L. and {van Gelder}, M.~L. and {Ray}, T.~P. and {Beuther}, H. and {Caratti o Garatti}, A. and {Brunken}, N.~G.~C. and {Chen}, Y. and {Devaraj}, R. and {Geers}, V.~C. and {Gieser}, C. and {Greene}, T.~P. and {Justtanont}, K. and {Le Gouellec}, V.~J.~M. and {Kavanagh}, P.~J. and {Klaassen}, P.~D. and {Janssen}, A.~G.~M. and {Navarro}, M.~G. and {Nazari}, P. and {Notsu}, S. and {Perotti}, G. and {Ressler}, M.~E. and {Reyes}, S.~D. and {Sellek}, A.~D. and {Tabone}, B. and {Tap}, C. and {Theijssen}, N.~C.~M.~A. and {Colina}, L. and {G{\"u}del}, M. and {Henning}, Th. and {Lagage}, P. -O. and {{\"O}stlin}, G. and {Vandenbussche}, B. and {Wright}, G.~S.},
        title = "{JWST Observations of Young protoStars (JOYS): Overview of program and early results}",
      journal = {\aap},
         year = 2025,
        month = jul,
       volume = {699},
          eid = {A361},
        pages = {A361},
          doi = {10.1051/0004-6361/202554444},
archivePrefix = {arXiv},
       eprint = {2505.08002},
 primaryClass = {astro-ph.GA},
       adsurl = {https://ui.adsabs.harvard.edu/abs/2025A&A...699A.361V}
}

@ARTICLE{Federman2024_IPA,
       author = {{Federman}, Samuel A. and {Megeath}, S. Thomas and {Rubinstein}, Adam E. and {Gutermuth}, Robert and {Narang}, Mayank and {Tyagi}, Himanshu and {Manoj}, P. and {Anglada}, Guillem and {Atnagulov}, Prabhani and {Beuther}, Henrik and {Bourke}, Tyler L. and {Brunken}, Nashanty and {Caratti o Garatti}, Alessio and {Evans}, Neal J. and {Fischer}, William J. and {Furlan}, Elise and {Green}, Joel D. and {Habel}, Nolan and {Hartmann}, Lee and {Karnath}, Nicole and {Klaassen}, Pamela and {Linz}, Hendrik and {Looney}, Leslie W. and {Osorio}, Mayra and {Muzerolle Page}, James and {Nazari}, Pooneh and {Pokhrel}, Riwaj and {Rahatgaonkar}, Rohan and {Rocha}, Will R.~M. and {Sheehan}, Patrick and {Slavicinska}, Katerina and {Stanke}, Thomas and {Stutz}, Amelia M. and {Tobin}, John J. and {Tychoniec}, Lukasz and {Van Dishoeck}, Ewine F. and {Watson}, Dan M. and {Wolk}, Scott and {Yang}, Yao-Lun},
        title = "{Investigating Protostellar Accretion-driven Outflows across the Mass Spectrum: JWST NIRSpec Integral Field Unit 3{\textendash}5 {\ensuremath{\mu}}m Spectral Mapping of Five Young Protostars}",
      journal = {\apj},
         year = 2024,
        month = may,
       volume = {966},
       number = {1},
          eid = {41},
        pages = {41},
          doi = {10.3847/1538-4357/ad2fa0},
archivePrefix = {arXiv},
       eprint = {2310.03803},
 primaryClass = {astro-ph.SR},
       adsurl = {https://ui.adsabs.harvard.edu/abs/2024ApJ...966...41F}
}

@ARTICLE{vanGelder2024_SO2,
       author = {{van Gelder}, M.~L. and {Ressler}, M.~E. and {van Dishoeck}, E.~F. and {Nazari}, P. and {Tabone}, B. and {Black}, J.~H. and {Tychoniec}, {\L}. and {Francis}, L. and {Barsony}, M. and {Beuther}, H. and {Caratti o Garatti}, A. and {Chen}, Y. and {Gieser}, C. and {le Gouellec}, V.~J.~M. and {Kavanagh}, P.~J. and {Klaassen}, P.~D. and {Lew}, B.~W.~P. and {Linnartz}, H. and {Majumdar}, L. and {Perotti}, G. and {Rocha}, W.~R.~M.},
        title = "{JOYS+: Mid-infrared detection of gas-phase SO$_{2}$ emission in a low-mass protostar. The case of NGC 1333 IRAS 2A: Hot core or accretion shock?}",
      journal = {\aap},
         year = 2024,
        month = feb,
       volume = {682},
          eid = {A78},
        pages = {A78},
          doi = {10.1051/0004-6361/202348118},
archivePrefix = {arXiv},
       eprint = {2311.17161},
 primaryClass = {astro-ph.SR},
       adsurl = {https://ui.adsabs.harvard.edu/abs/2024A&A...682A..78V}
}

@ARTICLE{vanGelder2024_gas,
       author = {{van Gelder}, M.~L. and {Francis}, L. and {van Dishoeck}, E.~F. and {Tychoniec}, {\L}. and {Ray}, T.~P. and {Beuther}, H. and {Caratti o Garatti}, A. and {Chen}, Y. and {Devaraj}, R. and {Gieser}, C. and {Justtanont}, K. and {Kavanagh}, P.~J. and {Nazari}, P. and {Reyes}, S. and {Rocha}, W.~R.~M. and {Slavicinska}, K. and {G{\"u}del}, M. and {Henning}, Th. and {Lagage}, P. -O. and {Wright}, G.},
        title = "{JWST Observations of Young protoStars (JOYS): Overview of gaseous molecular emission and absorption in low-mass protostars}",
        journal = {\aap},
         year = 2024,
        month = dec,
       volume = {692},
          eid = {A197},
        pages = {A197},
          doi = {10.1051/0004-6361/202451967},
archivePrefix = {arXiv},
       eprint = {2410.01636},
 primaryClass = {astro-ph.SR},
       adsurl = {https://ui.adsabs.harvard.edu/abs/2024A&A...692A.197V}
}

@ARTICLE{Law2023,
       author = {{Law}, David R. and {E. Morrison}, Jane and {Argyriou}, Ioannis and {Patapis}, Polychronis and {{\'A}lvarez-M{\'a}rquez}, J. and {Labiano}, Alvaro and {Vandenbussche}, Bart},
        title = "{A 3D Drizzle Algorithm for JWST and Practical Application to the MIRI Medium Resolution Spectrometer}",
      journal = {\aj},
         year = 2023,
        month = aug,
       volume = {166},
       number = {2},
          eid = {45},
        pages = {45},
          doi = {10.3847/1538-3881/acdddc},
archivePrefix = {arXiv},
       eprint = {2306.05520},
 primaryClass = {astro-ph.IM},
       adsurl = {https://ui.adsabs.harvard.edu/abs/2023AJ....166...45L}
}

@ARTICLE{Tabone2023,
       author = {{Tabone}, B. and {Bettoni}, G. and {van Dishoeck}, E.~F. and {Arabhavi}, A.~M. and {Grant}, S. and {Gasman}, D. and {Henning}, Th. and {Kamp}, I. and {G{\"u}del}, M. and {Lagage}, P.~O. and {Ray}, T. and {Vandenbussche}, B. and {Abergel}, A. and {Absil}, O. and {Argyriou}, I. and {Barrado}, D. and {Boccaletti}, A. and {Bouwman}, J. and {Caratti o Garatti}, A. and {Geers}, V. and {Glauser}, A.~M. and {Justannont}, K. and {Lahuis}, F. and {Mueller}, M. and {Nehm{\'e}}, C. and {Olofsson}, G. and {Pantin}, E. and {Scheithauer}, S. and {Waelkens}, C. and {Waters}, L.~B.~F.~M. and {Black}, J.~H. and {Christiaens}, V. and {Guadarrama}, R. and {Morales-Calder{\'o}n}, M. and {Jang}, H. and {Kanwar}, J. and {Pawellek}, N. and {Perotti}, G. and {Perrin}, A. and {Rodgers-Lee}, D. and {Samland}, M. and {Schreiber}, J. and {Schwarz}, K. and {Colina}, L. and {{\"O}stlin}, G. and {Wright}, G.},
        title = "{A rich hydrocarbon chemistry and high C to O ratio in the inner disk around a very low-mass star}",
      journal = {Nature Astronomy},
         year = 2023,
        month = jul,
       volume = {7},
        pages = {805-814},
          doi = {10.1038/s41550-023-01965-3},
archivePrefix = {arXiv},
       eprint = {2304.05954},
 primaryClass = {astro-ph.EP},
       adsurl = {https://ui.adsabs.harvard.edu/abs/2023NatAs...7..805T}
}

@ARTICLE{Francis2024,
       author = {{Francis}, L. and {van Gelder}, M.~L. and {van Dishoeck}, E.~F. and {Gieser}, C. and {Beuther}, H. and {Tychoniec}, L. and {Perotti}, G. and {Caratti o Garatti}, A. and {Kavanagh}, P.~J. and {Ray}, T. and {Klaassen}, P. and {Justtanont}, K. and {Linnartz}, H. and {Rocha}, W.~R.~M. and {Slavicinska}, K. and {G{\"u}del}, M. and {Henning}, T. and {Lagage}, P. -O. and {{\"O}stlin}, G.},
        title = "{JOYS: MIRI/MRS spectroscopy of gas-phase molecules from the high-mass star-forming region IRAS 23385+6053}",
      journal = {\aap},
         year = 2024,
        month = mar,
       volume = {683},
          eid = {A249},
        pages = {A249},
          doi = {10.1051/0004-6361/202348105},
archivePrefix = {arXiv},
       eprint = {2401.06880},
 primaryClass = {astro-ph.GA},
       adsurl = {https://ui.adsabs.harvard.edu/abs/2024A&A...683A.249F}
}

@ARTICLE{Boogert2008,
       author = {{Boogert}, A.~C.~A. and {Pontoppidan}, K.~M. and {Knez}, C. and {Lahuis}, F. and {Kessler-Silacci}, J. and {van Dishoeck}, E.~F. and {Blake}, G.~A. and {Augereau}, J. -C. and {Bisschop}, S.~E. and {Bottinelli}, S. and {Brooke}, T.~Y. and {Brown}, J. and {Crapsi}, A. and {Evans}, II, N.~J. and {Fraser}, H.~J. and {Geers}, V. and {Huard}, T.~L. and {J{\o}rgensen}, J.~K. and {{\"O}berg}, K.~I. and {Allen}, L.~E. and {Harvey}, P.~M. and {Koerner}, D.~W. and {Mundy}, L.~G. and {Padgett}, D.~L. and {Sargent}, A.~I. and {Stapelfeldt}, K.~R.},
        title = "{The c2d Spitzer Spectroscopic Survey of Ices around Low-Mass Young Stellar Objects. I. H$_{2}$O and the 5-8 {\ensuremath{\mu}}m Bands}",
      journal = {\apj},
         year = 2008,
        month = may,
       volume = {678},
       number = {2},
        pages = {985-1004},
          doi = {10.1086/533425},
archivePrefix = {arXiv},
       eprint = {0801.1167},
 primaryClass = {astro-ph},
       adsurl = {https://ui.adsabs.harvard.edu/abs/2008ApJ...678..985B}
}

@ARTICLE{Gibb2004,
       author = {{Gibb}, E.~L. and {Whittet}, D.~C.~B. and {Boogert}, A.~C.~A. and {Tielens}, A.~G.~G.~M.},
        title = "{Interstellar Ice: The Infrared Space Observatory Legacy}",
      journal = {\apjs},
         year = 2004,
        month = mar,
       volume = {151},
       number = {1},
        pages = {35-73},
          doi = {10.1086/381182},
       adsurl = {https://ui.adsabs.harvard.edu/abs/2004ApJS..151...35G}
}

@ARTICLE{Rocha2025,
       author = {{Rocha}, W.~R.~M. and {McClure}, M.~K. and {Sturm}, J.~A. and {Beck}, T.~L. and {Smith}, Z.~L. and {Dickinson}, H. and {Sun}, F. and {Egami}, E. and {Boogert}, A.~C.~A. and {Fraser}, H.~J. and {Dartois}, E. and {Jimenez-Serra}, I. and {Noble}, J.~A. and {Bergner}, J. and {Caselli}, P. and {Charnley}, S.~B. and {Chiar}, J. and {Chu}, L. and {Cooke}, I. and {Crouzet}, N. and {van Dishoeck}, E.~F. and {Drozdovskaya}, M.~N. and {Garrod}, R. and {Harsono}, D. and {Ioppolo}, S. and {Jin}, M. and {J{\o}rgensen}, J.~K. and {Lamberts}, T. and {Lis}, D.~C. and {Melnick}, G.~J. and {McGuire}, B.~A. and {{\"O}berg}, K.~I. and {Palumbo}, M.~E. and {Pendleton}, Y.~J. and {Perotti}, G. and {Qasim}, D. and {Shope}, B. and {Urso}, R.~G. and {Viti}, S. and {Linnartz}, H.},
        title = "{Ice inventory towards the protostar Ced 110 IRS4 observed with the James Webb Space Telescope: Results from the Early Release Science Ice Age program}",
      journal = {\aap},
         year = 2025,
        month = jan,
       volume = {693},
          eid = {A288},
        pages = {A288},
          doi = {10.1051/0004-6361/202451505},
archivePrefix = {arXiv},
       eprint = {2411.19651},
 primaryClass = {astro-ph.SR},
       adsurl = {https://ui.adsabs.harvard.edu/abs/2025A&A...693A.288R}
}

@ARTICLE{Turner2026,
       author = {{Turner}, Andrew M. and {Yang}, Yao-Lun and {Gross}, Rachel and {Sakai}, Nami and {Kaiser}, Ralf I.},
        title = "{A Chemistry-first Centered Icy Chemical Inventory of Protostellar Sources with JWST}",
      journal = {\apj},
         year = 2026,
        month = mar,
       volume = {1000},
       number = {1},
          eid = {17},
        pages = {17},
          doi = {10.3847/1538-4357/ae3743},
archivePrefix = {arXiv},
       eprint = {2602.05383},
 primaryClass = {astro-ph.GA},
       adsurl = {https://ui.adsabs.harvard.edu/abs/2026ApJ..1000...17T}
}

@ARTICLE{Gieser2026,
       author = {{Gieser}, C. and {Rocha}, W.~R.~M. and {Chen}, Y. and {Slavicinska}, K. and {van Dishoeck}, E.~F. and {Nazari}, P. and {Brunken}, N.~G.~C. and {Francis}, L. and {Beuther}, H. and {Reyes-Reyes}, S. and {Caratti o Garatti}, A. and {Klaassen}, P.~D. and {Vorster}, J.~M. and {Navarro}, M.~G.},
        title = "{JOYS: Linking the molecular ice and gas phase composition toward the high-mass hot core IRAS 18089─1732}",
      journal = {\aap},
         year = 2026,
        month = may,
       volume = {709},
          eid = {A235},
        pages = {A235},
          doi = {10.1051/0004-6361/202558615},
archivePrefix = {arXiv},
       eprint = {2603.22007},
 primaryClass = {astro-ph.GA},
       adsurl = {https://ui.adsabs.harvard.edu/abs/2026A&A...709A.235G}
}

@ARTICLE{Rayalacheruvu2025,
       author = {{Rayalacheruvu}, Prathap and {Majumdar}, Liton and {Rocha}, W.~R.~M. and {Ressler}, Michael E. and {Giri}, Pabitra Ranjan and {Maitrey}, S. and {Willacy}, Karen and {Lis}, D.~C. and {Chen}, Yuan and {Klaassen}, P.~D.},
        title = "{Expanding the Ice Inventory of NGC 1333 IRAS 2A with INDRA Using JWST Observations: Tracing Organic Refractories and Beyond}",
      journal = {\apjs},
         year = 2025,
        month = dec,
       volume = {281},
       number = {2},
          eid = {51},
        pages = {51},
          doi = {10.3847/1538-4365/ae112f},
archivePrefix = {arXiv},
       eprint = {2506.15358},
 primaryClass = {astro-ph.GA},
       adsurl = {https://ui.adsabs.harvard.edu/abs/2025ApJS..281...51R}
}

@software{Dominik2021_optool,
       author = {{Dominik}, Carsten and {Min}, Michiel and {Tazaki}, Ryo},
        title = "{OpTool: Command-line driven tool for creating complex dust opacities}",
 howpublished = {Astrophysics Source Code Library, record ascl:2104.010},
         year = 2021,
        month = apr,
          eid = {ascl:2104.010},
archivePrefix = {ascl},
       eprint = {2104.010},
       adsurl = {https://ui.adsabs.harvard.edu/abs/2021ascl.soft04010D}
}

@MISC{Regan2023,
       author = {{Regan}, Michael},
        title = "{Detection and Flagging of Showers and Snowballs in JWST}",
 howpublished = {Technical Report JWST-STScI-008545},
         year = 2023,
        month = sep,
        pages = {8545},
       adsurl = {https://ui.adsabs.harvard.edu/abs/2023jwst.rept.8545R}
}

@ARTICLE{Boogert2015,
       author = {{Boogert}, A.~C. Adwin and {Gerakines}, Perry A. and {Whittet}, Douglas C.~B.},
        title = "{Observations of the icy universe.}",
      journal = {\araa},
         year = 2015,
        month = aug,
       volume = {53},
        pages = {541-581},
          doi = {10.1146/annurev-astro-082214-122348},
archivePrefix = {arXiv},
       eprint = {1501.05317},
 primaryClass = {astro-ph.GA},
       adsurl = {https://ui.adsabs.harvard.edu/abs/2015ARA&A..53..541B}
}

@ARTICLE{TvS2018,
       author = {{Terwisscha van Scheltinga}, J. and {Ligterink}, N.~F.~W. and {Boogert}, A.~C.~A. and {van Dishoeck}, E.~F. and {Linnartz}, H.},
        title = "{Infrared spectra of complex organic molecules in astronomically relevant ice matrices. I. Acetaldehyde, ethanol, and dimethyl ether}",
      journal = {\aap},
         year = 2018,
        month = mar,
       volume = {611},
          eid = {A35},
        pages = {A35},
          doi = {10.1051/0004-6361/201731998},
archivePrefix = {arXiv},
       eprint = {1712.04796},
 primaryClass = {astro-ph.SR},
       adsurl = {https://ui.adsabs.harvard.edu/abs/2018A&A...611A..35T}
}

@ARTICLE{Rachid2020,
       author = {{Rachid}, M.~G. and {Terwisscha van Scheltinga}, J. and {Koletzki}, D. and {Linnartz}, H.},
        title = "{Infrared spectra of complex organic molecules in astronomically relevant ice mixtures. II. Acetone}",
      journal = {\aap},
         year = 2020,
        month = jul,
       volume = {639},
          eid = {A4},
        pages = {A4},
          doi = {10.1051/0004-6361/202037497},
       adsurl = {https://ui.adsabs.harvard.edu/abs/2020A&A...639A...4R}
}

@ARTICLE{Nazari2022_NCOM,
       author = {{Nazari}, P. and {Meijerhof}, J.~D. and {van Gelder}, M.~L. and {Ahmadi}, A. and {van Dishoeck}, E.~F. and {Tabone}, B. and {Langeroodi}, D. and {Ligterink}, N.~F.~W. and {Jaspers}, J. and {Beltr{\'a}n}, M.~T. and {Fuller}, G.~A. and {S{\'a}nchez-Monge}, {\'A}. and {Schilke}, P.},
        title = "{N-bearing complex organics toward high-mass protostars. Constant ratios pointing to formation in similar pre-stellar conditions across a large mass range}",
      journal = {\aap},
         year = 2022,
        month = dec,
       volume = {668},
          eid = {A109},
        pages = {A109},
          doi = {10.1051/0004-6361/202243788},
archivePrefix = {arXiv},
       eprint = {2208.11128},
 primaryClass = {astro-ph.GA},
       adsurl = {https://ui.adsabs.harvard.edu/abs/2022A&A...668A.109N}
}

@ARTICLE{Schutte1999,
       author = {{Schutte}, W.~A. and {Boogert}, A.~C.~A. and {Tielens}, A.~G.~G.~M. and {Whittet}, D.~C.~B. and {Gerakines}, P.~A. and {Chiar}, J.~E. and {Ehrenfreund}, P. and {Greenberg}, J.~M. and {van Dishoeck}, E.~F. and {de Graauw}, Th.},
        title = "{Weak ice absorption features at 7.24 and 7.41 MU M in the spectrum of the obscured young stellar object W 33A}",
      journal = {\aap},
         year = 1999,
        month = mar,
       volume = {343},
        pages = {966-976},
       adsurl = {https://ui.adsabs.harvard.edu/abs/1999A&A...343..966S}
}

@ARTICLE{Oberg2011,
       author = {{{\"O}berg}, Karin I. and {Boogert}, A.~C. Adwin and {Pontoppidan}, Klaus M. and {van den Broek}, Saskia and {van Dishoeck}, Ewine F. and {Bottinelli}, Sandrine and {Blake}, Geoffrey A. and {Evans}, Neal J., II},
        title = "{The Spitzer Ice Legacy: Ice Evolution from Cores to Protostars}",
      journal = {\apj},
         year = 2011,
        month = oct,
       volume = {740},
       number = {2},
          eid = {109},
        pages = {109},
          doi = {10.1088/0004-637X/740/2/109},
archivePrefix = {arXiv},
       eprint = {1107.5825},
 primaryClass = {astro-ph.GA},
       adsurl = {https://ui.adsabs.harvard.edu/abs/2011ApJ...740..109O}
}

@ARTICLE{Nazari2024_CH3CNice,
       author = {{Nazari}, P. and {Rocha}, W.~R.~M. and {Rubinstein}, A.~E. and {Slavicinska}, K. and {Rachid}, M.~G. and {van Dishoeck}, E.~F. and {Megeath}, S.~T. and {Gutermuth}, R. and {Tyagi}, H. and {Brunken}, N. and {Narang}, M. and {Manoj}, P. and {Watson}, D.~M. and {Evans}, N.~J. and {Federman}, S. and {Muzerolle Page}, J. and {Anglada}, G. and {Beuther}, H. and {Klaassen}, P. and {Looney}, L.~W. and {Osorio}, M. and {Stanke}, T. and {Yang}, Y. -L.},
        title = "{Hunting for complex cyanides in protostellar ices with the JWST. A tentative detection of CH$_{3}$CN and C$_{2}$H$_{5}$CN}",
      journal = {\aap},
         year = 2024,
        month = jun,
       volume = {686},
          eid = {A71},
        pages = {A71},
          doi = {10.1051/0004-6361/202348695},
archivePrefix = {arXiv},
       eprint = {2401.07901},
 primaryClass = {astro-ph.GA},
       adsurl = {https://ui.adsabs.harvard.edu/abs/2024A&A...686A..71N}
}

@ARTICLE{Chen2023,
       author = {{Chen}, Y. and {van Gelder}, M.~L. and {Nazari}, P. and {Brogan}, C.~L. and {van Dishoeck}, E.~F. and {Linnartz}, H. and {J{\o}rgensen}, J.~K. and {Hunter}, T.~R. and {Wilkins}, O.~H. and {Blake}, G.~A. and {Caselli}, P. and {Chuang}, K. -J. and {Codella}, C. and {Cooke}, I. and {Drozdovskaya}, M.~N. and {Garrod}, R.~T. and {Ioppolo}, S. and {Jin}, M. and {Kulterer}, B.~M. and {Ligterink}, N.~F.~W. and {Lipnicky}, A. and {Loomis}, R. and {Rachid}, M.~G. and {Spezzano}, S. and {McGuire}, B.~A.},
        title = "{CoCCoA: Complex Chemistry in hot Cores with ALMA. Selected oxygen-bearing species}",
      journal = {\aap},
         year = 2023,
        month = oct,
       volume = {678},
          eid = {A137},
        pages = {A137},
          doi = {10.1051/0004-6361/202346491},
archivePrefix = {arXiv},
       eprint = {2308.02688},
 primaryClass = {astro-ph.GA},
       adsurl = {https://ui.adsabs.harvard.edu/abs/2023A&A...678A.137C}
}

@article{Rocha2024,
	author = {{Rocha}, W. R. M. and {van Dishoeck, E. F.} and {Ressler, M. E.} and {van Gelder, M. L.} and {Slavicinska, K.} and {Brunken, N. G. C.} and {Linnartz, H.} and {Ray, T. P.} and {Beuther, H.} and {Caratti o Garatti, A.} and {Geers, V.} and {Kavanagh, P. J.} and {Klaassen, P. D.} and {Justtanont, K.} and {Chen, Y.} and {Francis, L.} and {Gieser, C.} and {Perotti, G.} and {Tychoniec, Ł.} and {Barsony, M.} and {Majumdar, L.} and {le Gouellec, V. J. M.} and {Chu, L. E. U.} and {Lew, B. W. P.} and {Henning, Th.} and {Wright, G.}},
	title = {JWST Observations of Young protoStars (JOYS+): Detecting icy complex organic molecules and ions - I. CH4, SO2, HCOO−, OCN−, H2CO, HCOOH, CH3CH2OH, CH3CHO, CH3OCHO, and CH3COOH},
	DOI= "10.1051/0004-6361/202348427",
	url= "https://doi.org/10.1051/0004-6361/202348427",
	journal = {\aap},
	year = 2024,
	volume = 683,
	pages = "A124",
}

@ARTICLE{Chen2024,
       author = {{Chen}, Y. and {Rocha}, W.~R.~M. and {van Dishoeck}, E.~F. and {van Gelder}, M.~L. and {Nazari}, P. and {Slavicinska}, K. and {Francis}, L. and {Tabone}, B. and {Ressler}, M.~E. and {Klaassen}, P.~D. and {Beuther}, H. and {Boogert}, A.~C.~A. and {Gieser}, C. and {Kavanagh}, P.~J. and {Perotti}, G. and {Le Gouellec}, V.~J.~M. and {Majumdar}, L. and {G{\"u}del}, M. and {Henning}, Th.},
        title = "{JOYS+: The link between the ice and gas of complex organic molecules: Comparing JWST and ALMA data of two low-mass protostars}",
      journal = {\aap},
         year = 2024,
        month = oct,
       volume = {690},
          eid = {A205},
        pages = {A205},
          doi = {10.1051/0004-6361/202450706},
       adsurl = {https://ui.adsabs.harvard.edu/abs/2024A&A...690A.205C}
}

@ARTICLE{Fedoseev2022,
       author = {{Fedoseev}, Gleb and {Qasim}, Danna and {Chuang}, Ko-Ju and {Ioppolo}, Sergio and {Lamberts}, Thanja and {van Dishoeck}, Ewine F. and {Linnartz}, Harold},
        title = "{Hydrogenation of Accreting C Atoms and CO Molecules-Simulating Ketene and Acetaldehyde Formation Under Dark and Translucent Cloud Conditions}",
      journal = {\apj},
         year = 2022,
        month = jan,
       volume = {924},
       number = {2},
          eid = {110},
        pages = {110},
          doi = {10.3847/1538-4357/ac3834},
archivePrefix = {arXiv},
       eprint = {2111.08548},
 primaryClass = {astro-ph.GA},
       adsurl = {https://ui.adsabs.harvard.edu/abs/2022ApJ...924..110F}
}

@ARTICLE{Watanabe2002,
       author = {{Watanabe}, Naoki and {Kouchi}, Akira},
        title = "{Efficient Formation of Formaldehyde and Methanol by the Addition of Hydrogen Atoms to CO in H$_{2}$O-CO Ice at 10 K}",
      journal = {\apjl},
         year = 2002,
        month = jun,
       volume = {571},
       number = {2},
        pages = {L173-L176},
          doi = {10.1086/341412},
       adsurl = {https://ui.adsabs.harvard.edu/abs/2002ApJ...571L.173W}
}

@ARTICLE{Fuchs2009,
       author = {{Fuchs}, G.~W. and {Cuppen}, H.~M. and {Ioppolo}, S. and {Romanzin}, C. and {Bisschop}, S.~E. and {Andersson}, S. and {van Dishoeck}, E.~F. and {Linnartz}, H.},
        title = "{Hydrogenation reactions in interstellar CO ice analogues. A combined experimental/theoretical approach}",
      journal = {\aap},
         year = 2009,
        month = oct,
       volume = {505},
       number = {2},
        pages = {629-639},
          doi = {10.1051/0004-6361/200810784},
       adsurl = {https://ui.adsabs.harvard.edu/abs/2009A&A...505..629F}
}

@ARTICLE{Simons2020,
       author = {{Simons}, M.~A.~J. and {Lamberts}, T. and {Cuppen}, H.~M.},
        title = "{Formation of COMs through CO hydrogenation on interstellar grains}",
      journal = {\aap},
         year = 2020,
        month = feb,
       volume = {634},
          eid = {A52},
        pages = {A52},
          doi = {10.1051/0004-6361/201936522},
archivePrefix = {arXiv},
       eprint = {2001.04895},
 primaryClass = {astro-ph.SR},
       adsurl = {https://ui.adsabs.harvard.edu/abs/2020A&A...634A..52S}
}

@ARTICLE{Slavicinska2025_NH4SH,
       author = {{Slavicinska}, K. and {Boogert}, A.~C.~A. and {Tychoniec}, {\L}. and {van Dishoeck}, E.~F. and {van Gelder}, M.~L. and {Navarro}, M.~G. and {Santos}, J.~C. and {Klaassen}, P.~D. and {Kavanagh}, P.~J. and {Chuang}, K. -J.},
        title = "{Ammonium hydrosulfide (NH$_{4}$SH) as a potentially significant sulfur sink in interstellar ices}",
      journal = {\aap},
         year = 2025,
        month = jan,
       volume = {693},
          eid = {A146},
        pages = {A146},
          doi = {10.1051/0004-6361/202451383},
archivePrefix = {arXiv},
       eprint = {2410.02860},
 primaryClass = {astro-ph.GA},
       adsurl = {https://ui.adsabs.harvard.edu/abs/2025A&A...693A.146S}
}

@ARTICLE{Yarnall2022,
       author = {{Yarnall}, Yukiko Y. and {Hudson}, Reggie L.},
        title = "{A New Method for Measuring Infrared Band Strengths in H$_{2}$O Ices: First Results for OCS, H$_{2}$S, and SO$_{2}$}",
      journal = {\apjl},
         year = 2022,
        month = may,
       volume = {931},
       number = {1},
          eid = {L4},
        pages = {L4},
          doi = {10.3847/2041-8213/ac6b32},
       adsurl = {https://ui.adsabs.harvard.edu/abs/2022ApJ...931L...4Y}
}

@ARTICLE{Hudson2022,
       author = {{Hudson}, Reggie L. and {Gerakines}, Perry A. and {Yarnall}, Yukiko Y.},
        title = "{Ammonia Ices Revisited: New IR Intensities and Optical Constants for Solid NH$_{3}$}",
      journal = {\apj},
         year = 2022,
        month = feb,
       volume = {925},
       number = {2},
          eid = {156},
        pages = {156},
          doi = {10.3847/1538-4357/ac3e74},
       adsurl = {https://ui.adsabs.harvard.edu/abs/2022ApJ...925..156H}
}

@ARTICLE{Yang2022,
       author = {{Yang}, Yao-Lun and {Green}, Joel D. and {Pontoppidan}, Klaus M. and {Bergner}, Jennifer B. and {Cleeves}, L. Ilsedore and {Evans}, II, Neal J. and {Garrod}, Robin T. and {Jin}, Miwha and {Kim}, Chul Hwan and {Kim}, Jaeyeong and {Lee}, Jeong-Eun and {Sakai}, Nami and {Shingledecker}, Christopher N. and {Shope}, Brielle and {Tobin}, John J. and {van Dishoeck}, Ewine F.},
        title = "{CORINOS. I. JWST/MIRI Spectroscopy and Imaging of a Class 0 Protostar IRAS 15398{\textendash}3359}",
      journal = {\apjl},
         year = 2022,
        month = dec,
       volume = {941},
       number = {1},
          eid = {L13},
        pages = {L13},
          doi = {10.3847/2041-8213/aca289},
archivePrefix = {arXiv},
       eprint = {2208.10673},
 primaryClass = {astro-ph.SR},
       adsurl = {https://ui.adsabs.harvard.edu/abs/2022ApJ...941L..13Y}
}

@article{Nazari2025,
    title = {Overview of complex organic molecule observations in protostellar systems},
    journal = {Life Sciences in Space Research},
    year = {2025},
    issn = {2214-5524},
    doi = {https://doi.org/10.1016/j.lssr.2025.10.005},
    url = {https://www.sciencedirect.com/science/article/pii/S2214552425001154},
    author = {{Nazari}, Pooneh},
}

@software{lmfit2014,
       author = {{Newville}, Matthew and {Stensitzki}, Till and {Allen}, Daniel B. and {Ingargiola}, Antonino},
        title = "{LMFIT: Non-Linear Least-Square Minimization and Curve-Fitting for Python}",
         year = 2014,
        month = sep,
          eid = {10.5281/zenodo.11813},
          doi = {10.5281/zenodo.11813},
      version = {0.8.0},
    publisher = {Zenodo},
       adsurl = {https://ui.adsabs.harvard.edu/abs/2014zndo.....11813N}
}

@Article{         Numpy2020,
 title         = {Array programming with {NumPy}},
 author        = {Charles R. Harris and K. Jarrod Millman and St{\'{e}}fan J.
                 van der Walt and Ralf Gommers and Pauli Virtanen and David
                 Cournapeau and Eric Wieser and Julian Taylor and Sebastian
                 Berg and Nathaniel J. Smith and Robert Kern and Matti Picus
                 and Stephan Hoyer and Marten H. van Kerkwijk and Matthew
                 Brett and Allan Haldane and Jaime Fern{\'{a}}ndez del
                 R{\'{i}}o and Mark Wiebe and Pearu Peterson and Pierre
                 G{\'{e}}rard-Marchant and Kevin Sheppard and Tyler Reddy and
                 Warren Weckesser and Hameer Abbasi and Christoph Gohlke and
                 Travis E. Oliphant},
 year          = {2020},
 month         = sep,
 journal       = {Nature},
 volume        = {585},
 number        = {7825},
 pages         = {357--362},
 doi           = {10.1038/s41586-020-2649-2},
 publisher     = {Springer Science and Business Media {LLC}},
 url           = {https://doi.org/10.1038/s41586-020-2649-2}
}

@Article{Matplotlib2007,
  Author    = {Hunter, J. D.},
  Title     = {Matplotlib: A 2D graphics environment},
  Journal   = {Computing in Science \& Engineering},
  Volume    = {9},
  Number    = {3},
  Pages     = {90--95},
  publisher = {IEEE COMPUTER SOC},
  doi       = {10.1109/MCSE.2007.55},
  year      = 2007
}

@ARTICLE{SciPy2020,
  author  = {Virtanen, Pauli and Gommers, Ralf and Oliphant, Travis E. and
            Haberland, Matt and Reddy, Tyler and Cournapeau, David and
            Burovski, Evgeni and Peterson, Pearu and Weckesser, Warren and
            Bright, Jonathan and {van der Walt}, St{\'e}fan J. and
            Brett, Matthew and Wilson, Joshua and Millman, K. Jarrod and
            Mayorov, Nikolay and Nelson, Andrew R. J. and Jones, Eric and
            Kern, Robert and Larson, Eric and Carey, C J and
            Polat, {\.I}lhan and Feng, Yu and Moore, Eric W. and
            {VanderPlas}, Jake and Laxalde, Denis and Perktold, Josef and
            Cimrman, Robert and Henriksen, Ian and Quintero, E. A. and
            Harris, Charles R. and Archibald, Anne M. and
            Ribeiro, Ant{\^o}nio H. and Pedregosa, Fabian and
            {van Mulbregt}, Paul and {SciPy 1.0 Contributors}},
  title   = {{{SciPy} 1.0: Fundamental Algorithms for Scientific
            Computing in Python}},
  journal = {Nature Methods},
  year    = {2020},
  volume  = {17},
  pages   = {261--272},
  adsurl  = {https://rdcu.be/b08Wh},
  doi     = {10.1038/s41592-019-0686-2},
}
\bibliographystyle{aasjournalv7}



\end{document}